\documentclass[%
    aip,
    pop,
    onecolumn,
    amsmath,amssymb,
    preprint,
]{revtex4-2} 

\usepackage{graphicx} 
\usepackage{dcolumn} 
\usepackage{bm} 
\usepackage{amsmath}

\usepackage[caption=false]{subfig}
\usepackage[title]{appendix}
\usepackage{hyperref}

\makeatletter
\def\@email#1#2{%
 \endgroup
 \patchcmd{\titleblock@produce}
  {\frontmatter@RRAPformat}
  {\frontmatter@RRAPformat{\produce@RRAP{*#1\href{mailto:#2}{#2}}}\frontmatter@RRAPformat}
  {}{}
}%
\makeatother
\begin{document}


\title[]{On the validity of quasilinear theory applied to the electron bump-on-tail instability}
\author{D.W. Crews}
\author{U. Shumlak}%
    \affiliation{Computational Plasma Dynamics Lab, Aerospace and Energetics Research Program,\\
    University of Washington, Seattle, Washington 98195-2400 USA.}

\date{\today}



    \begin{abstract}
    The accuracy of quasilinear theory applied to the electron bump-on-tail instability, a classic model problem,
    is explored with conservative high-order discontinuous Galerkin methods applied to both the
    quasilinear equations and to a direct simulation of the Vlasov-Poisson equations.
    The initial condition is chosen in the regime of beam parameters for which quasilinear theory should be applicable.
    Quasilinear diffusion is initially in good agreement with the direct simulation but later underestimates
    the turbulent momentum flux.
    The direct simulation corrects from quasilinear evolution by quenching the instability in a finite
    time and producing a robust state of oscillation.
    Flux enhancement above quasilinear levels occurs as the phase space eddy turnover time in the largest amplitude wavepackets becomes
    comparable to the transit time of resonant phase fluid through wavepacket potentials.
    In this regime eddies effectively turn over during wavepacket transit so that phase fluid predominantly disperses by
    eddy phase mixing rather than by randomly phased waves.
    The enhanced turbulent flux of resonant phase fluid leads in turn, through energy conservation,
    to an increase in non-resonant turbulent flux and thus to an enhanced heating of the main thermal body
    above quasilinear predictions.
    These findings shed light on the kinetic turbulence fluctuation spectrum and
    support the theory that collisionless momentum diffusion beyond the quasilinear approximation
    can be understood through the dynamics of phase space eddies (or clumps and granulations).
    \end{abstract}

    \maketitle

    \section{Introduction}\label{sec:intro}
    One-dimensional electrostatic turbulence is
    among the oldest model problems in plasma physics and stimulated the development of the
    quasilinear theory of plasma diffusion~\cite{drummond,vedenov_velikhov,krommes2015tutorial}.
    Quasilinear diffusion theories are used to predict
    the statistical properties of turbulent energy and the plasma distribution function in the saturated state
    of high-dimensional dynamic processes such as velocity space instability
    with a reduced low-dimensional model.
    Quasilinear methods have been widely applied to model turbulent diffusion in both physical and velocity space
    (\textit{e.g.}~gyrokinetic modeling~\cite{aspects}, rotating Couette flow~\cite{tobias_marston_2017},
    lower-hybrid drift instability~\cite{davidson_quasilinear}, firehose instability~\cite{quasilinear_firehose},
    electron interaction with whistlers~\cite{electron_diffusion}, etc.), often reproducing experimental phenomena
    beyond the theory's formal region of validity~\cite{surprising_success}.

    Quasilinear theory is concerned with the dynamics of macroscopic observables such as
    spatially-averaged quantities and the fluctuation power spectrum.
    It is applicable for weak turbulence, \textit{i.e.}~a broad spectrum of small-amplitude
    non-self-correlated linear waves in a homogeneous medium~\cite{cardy,diamond}.
    In order that phase interference, or wave beating, not lead to nonlinearity it is also
    necessary to suppose random phasing of the waves~\cite{davidson2012methods}.
    However the random phase approximation is somewhat fictional because the spectral width
    of unstable wavenumbers is finite.
    Interference of finite spectral width produces a spatial distribution of wave energy (wavepackets)
    even for random phases.
    For example the two-species numerical study in~\cite{Henri_2011} saw the late-time development of Langmuir cavitons
    from an initial ensemble of randomly-phased linear waves.
    The high amplitudes of wavepacket potentials lead to deviations from quasilinear predictions as
    linear waves grow from instability~\cite{vlad2004lagrangian}.

    Nevertheless quasilinear theory makes accurate predictions of the saturated state of instability
    despite the breakdown of the random phase approximation.
    This article demonstrates by a direct simulation
    of an electron bump-on-tail instability for which quasilinear theory should be applicable that weak nonlinearity due
    to spatial structure in the distribution of wave energy enhances the rate of turbulent flux above quasilinear levels.
    However the effect of this discrepancy is to reach a similar saturated state as the quasilinear prediction, but in a
    finite time rather than asymptotically with corresponding corrections to the transient linear growth rates and
    with significant fluctuations in the probability distribution function atop the quasilinear equilibrium.
    The success of quasilinear theory is in prediction of the greatest amplitude part of the field spectrum
    as this portion is due to linear growth.
    On the other hand, the quasilinear prediction does not capture some aspects of the weak kinetic turbulence spectrum such as
    plasma wave harmonics and the eddy turbulence of resonant particles,~\textit{i.e.} phase space granulations.

    Many numerical studies have been done comparing quasilinear analysis to solutions of the Vlasov equation using
    particle-in-cell (PIC) method~\cite{pic_firehose,pic_hall_thruster,pic_buneman}.
    Yet it is well-known that PIC methods are prone to errors due to statistically
    sampling the sensitive trajectories of the continuous
    distribution~\cite{cohen1989performance, brackbill2016energy, horky2017numerical}, so it is worthwhile to explore
    alternative kinetic modeling methods.
    In one such alternative the equation for the continuous distribution is solved by a demonstrably
    convergent and conservative finite element discretization of phase space,
    using for example discontinuous Galerkin methods~\cite{vlasov_dg,cheng2013study,juno2018discontinuous}.
    This work uses a high-order discontinuous Galerkin method to sufficiently resolve the detailed phase flow up to and
    past saturation of the kinetic instability.
    From this point of view the turbulent nature of resonant electrons, and the
    distinct behavior of resonant and non-resonant electrons, can be clearly seen.

    The structure of the article is a review of the theory of the bump-on-tail instability with an emphasis on the
    aspects necessary for a numerical study, followed by novel methods of discretization, and finally an analysis of the
    simulation results.
    Linear electrostatic theory is reviewed in Section~\ref{sec:review} in order to identify the kinetic eigenmodes
    used for the perturbation of the distribution function,
    followed by a review of the quasilinear theory of the bump-on-tail in Section~\ref{sec:quasilinear} with comments on
    the role of resonant and non-resonant parts of the distribution function and on the
    hierarchy of timescales required for validity of the approximation.
    The details of the discretization methods are discussed in Section~\ref{sec:discretization_details}.
    The simulation results are then explored in Section~\ref{sec:sim_results} by an analysis of phase space
    structures in the Vlasov-Poisson simulation and a comparison of quasilinear and Vlasov-Poisson levels of
    turbulent flux.

    \section{Linear theory of bump-on-tail instability}\label{sec:review}
    This section reviews electrostatic plasma instability with an emphasis on the linear response associated with
    unstable modes.
    One-dimensional collisionless electron dynamics in a uniform neutralizing background are governed by the
    Vlasov-Poisson equations~\cite{goldston},
    \begin{align}
        \frac{\partial f}{\partial t} &+ v\frac{\partial f}{\partial x} - E(x)\frac{\partial f}{\partial v} = 0,\label{eq:vlasov}\\
        \frac{dE}{dx} &= 1 - \int_{-\infty}^\infty f(x,v,t)dv\label{eq:gauss},
    \end{align}
    which describes the evolution of the probability distribution function $f(x,v,t)$ through phase space as influenced
    by the self-consistent electric field $E(x)$.
    Velocities are normalized to the thermal velocity $v_t$, lengths normalized to the Debye screening length $\lambda_D$,
    and electric field normalized to the thermal energy per Debye length.
    Static ions and collisionless trajectories are valid for timescales $\tau$ less than the ion plasma period
    and the electron-electron collision time, \textit{i.e.}~$\tau \ll \omega_{pi}^{-1}$
    and $\tau \ll \frac{\Lambda}{\log\Lambda} \omega_{pe}^{-1}$ where $\Lambda \gg 1$ is the inverse plasma parameter.
    This article simulates the Vlasov-Poisson system, and the result will be referred to as either the Vlasov-Poisson
    simulation or just the Vlasov simulation.

    \subsection{General linear theory of electrostatic instability}\label{sec:linear_theory}
    The linearized Vlasov-Poisson equations about a homogeneous equilibrium $f_0(v)$ are
    \begin{align}
        \frac{\partial f_1}{\partial t} &+ v\frac{\partial f_1}{\partial x} + \nabla\varphi\frac{\partial f_0}{\partial v} = 0,\\
        \nabla^2\varphi &= \int_{-\infty}^{\infty} f_1(x,v,t)dv
    \end{align}
    with $\varphi$ the potential.
    Following Landau~\cite{landau_damping}, spatial Fourier transform and one-sided temporal Fourier transform
    (or Laplace transform with $s\to -i\omega$) leads, if the initial condition is denoted $g(k,v)\equiv f_1(k,v,t=0)$,
    to a solution for electric potential,
    \begin{align}
        g^*(k,\zeta) &\equiv \int_{\mathcal{C}}\frac{g(k,v)}{\zeta - v}dv,\\
        \varphi(k,\zeta) &= -\frac{i}{k^3}\frac{g^*(k,\zeta)}{\varepsilon(k,\zeta)}\label{eq:complex_potential}
    \end{align}
    with $\zeta=\frac{\omega}{k}$ the complex phase velocity where the complex dielectric function is
    \begin{equation}\label{eq:dielectric}
        \varepsilon(k,\zeta) = 1 + \frac{1}{k^2}\int_{\mathcal{C}}\frac{1}{\zeta-v}\frac{\partial f_0}{\partial v}dv
    \end{equation}
    and $\mathcal{C}$ the Landau contour.
    Inversion of Eq.~\ref{eq:complex_potential} by the residue theorem suggests that the response to a
    general perturbation $g(k,\zeta)$ will include all solutions of $\varepsilon(k,\zeta)=0$.
    However initial conditions of the linear response form, with a single complex pole at an unstable frequency,
    excite only that mode.
    That is, suppose the initial condition is chosen as
    \begin{equation}\label{eq:special_ic}
        g(k,v) = \frac{1}{k^2}\frac{1}{\zeta_n - v}\frac{\partial f_0}{\partial v}
    \end{equation}
    where $\text{Im}(\zeta_n) > 0$ solves the dispersion relation in the upper half-plane,
    \textit{i.e.}~$\varepsilon(k,\zeta_n) = 0$.
    Then its transform along the Landau contour is the dielectric function with a pole at $\zeta=\zeta_n$,
    \begin{equation}\label{eq:special_ic_transform}
        g^*(k,\zeta) = -\frac{\varepsilon(k, \zeta)}{\zeta - \zeta_n}.
    \end{equation}
    The potential response is $\varphi(k,\zeta) = \frac{i}{k^3}\frac{1}{\zeta-\zeta_n}$, inverting to
    $\varphi(k,t) = \frac{1}{k^3}e^{-i\omega_n t}$.
    An initial condition in the form of Eq.~\ref{eq:special_ic} represents the discrete part of the
    spectrum of the linearized Vlasov-Poisson integral operator, as found by Case and
    van Kampen~\cite{case1959plasma,van1955theory}.
    Equation~\ref{eq:special_ic} is used in this work as the perturbation to excite the bump-on-tail instability.

    \subsection{Initialization of many discrete linear modes}\label{subsec:discrete_spectrum}
\begin{figure}[htp]
        \includegraphics[width=\columnwidth]{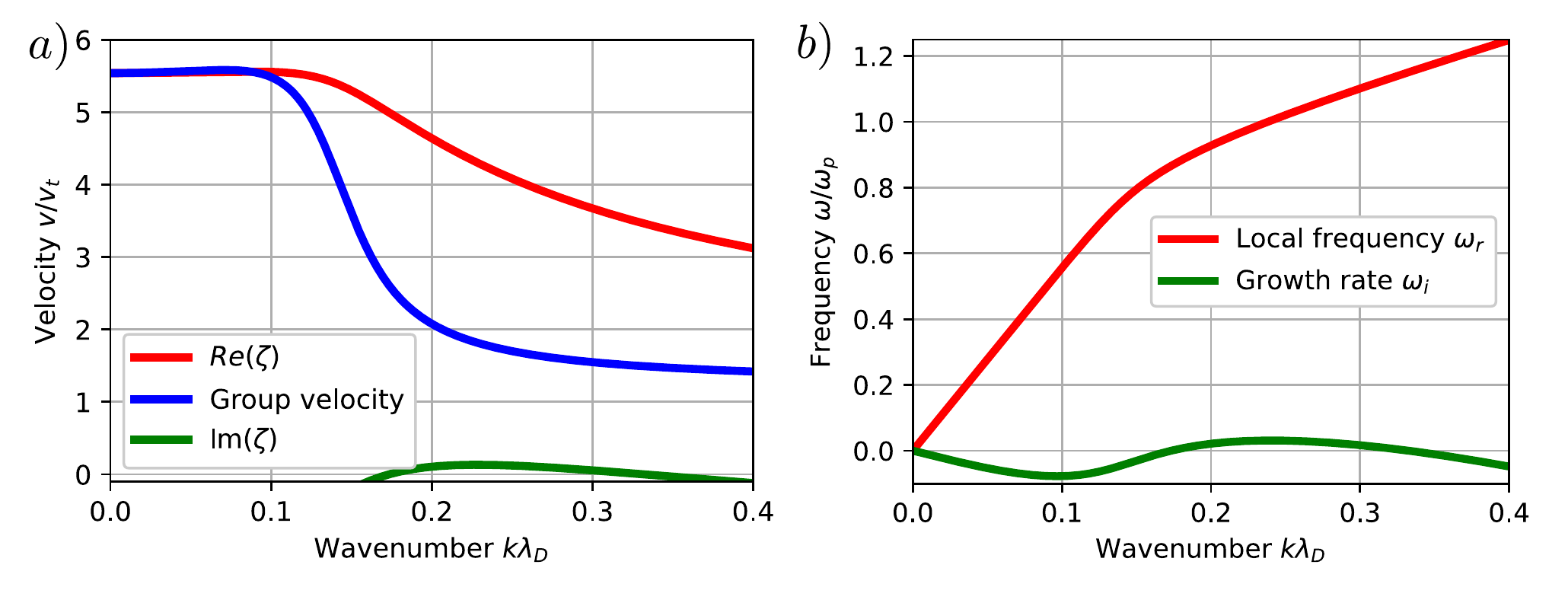}
        \caption{Bump-on-tail dispersion relation showing: a) wave velocities; and b) frequencies.
        The bump-on-tail distribution has an unstable branch of solutions with phase velocities
        bounded by the regions of $\frac{\partial f_0}{\partial v} > 0$ in the tail, in this case for $v\in (3,5)v_t$.
        The unstable solution is
            an electron-acoustic wave at the beam velocity for long wavelengths and an
            acoustic wave propagating at the bulk thermal velocity for short wavelengths.
            This unstable coupled acoustic wave is damped at both extremes and has
            maximum instability growth where the frequency $\omega\approx\omega_p$.
        }\label{fig:dispersion}
    \end{figure}

    An unstable distribution which should meet the applicability conditions of quasilinear theory
    consists of a hot drifting Maxwellian through a main thermal population
    \begin{equation}\label{eq:unstable_distribution}
    \begin{split}
        f_0(v) = \frac{1}{(1+\chi)\sqrt{2\pi}}\Big\{\frac{1}{v_{t0}}&\exp\Big(-\frac{v^2}{2v_{t0}^2}\Big)
    + \\&\frac{\chi}{v_{tb}}\exp\Big(-\frac{(v-v_b)^2}{2v_{tb}^2}\Big)\Big\}.
        \end{split}
    \end{equation}
    The bump-on-tail instability is the textbook example of a distribution with quasilinear evolution.
    According to Ref.~[\onlinecite{thorne2017modern}], reasoning by the Bohm-Gross dispersion relation,
    instability development should be in the weak turbulence regime given a bump fraction $\chi=0.05$,
    bump spread $v_{tb}=\chi^{1/3}v_b$, and beam velocity $v_b=5v_{t0}$.
    Velocities are then normalized to $v_{t0}=1$.
    Figure~\ref{fig:dispersion} shows the unstable branch of $\varepsilon(k,\zeta)=0$
    corresponding to the coupling of right-going oscillations of the main body and left-going
    oscillations on the drifting beam.

\begin{figure}[htp]
        \includegraphics[width=0.9\columnwidth]{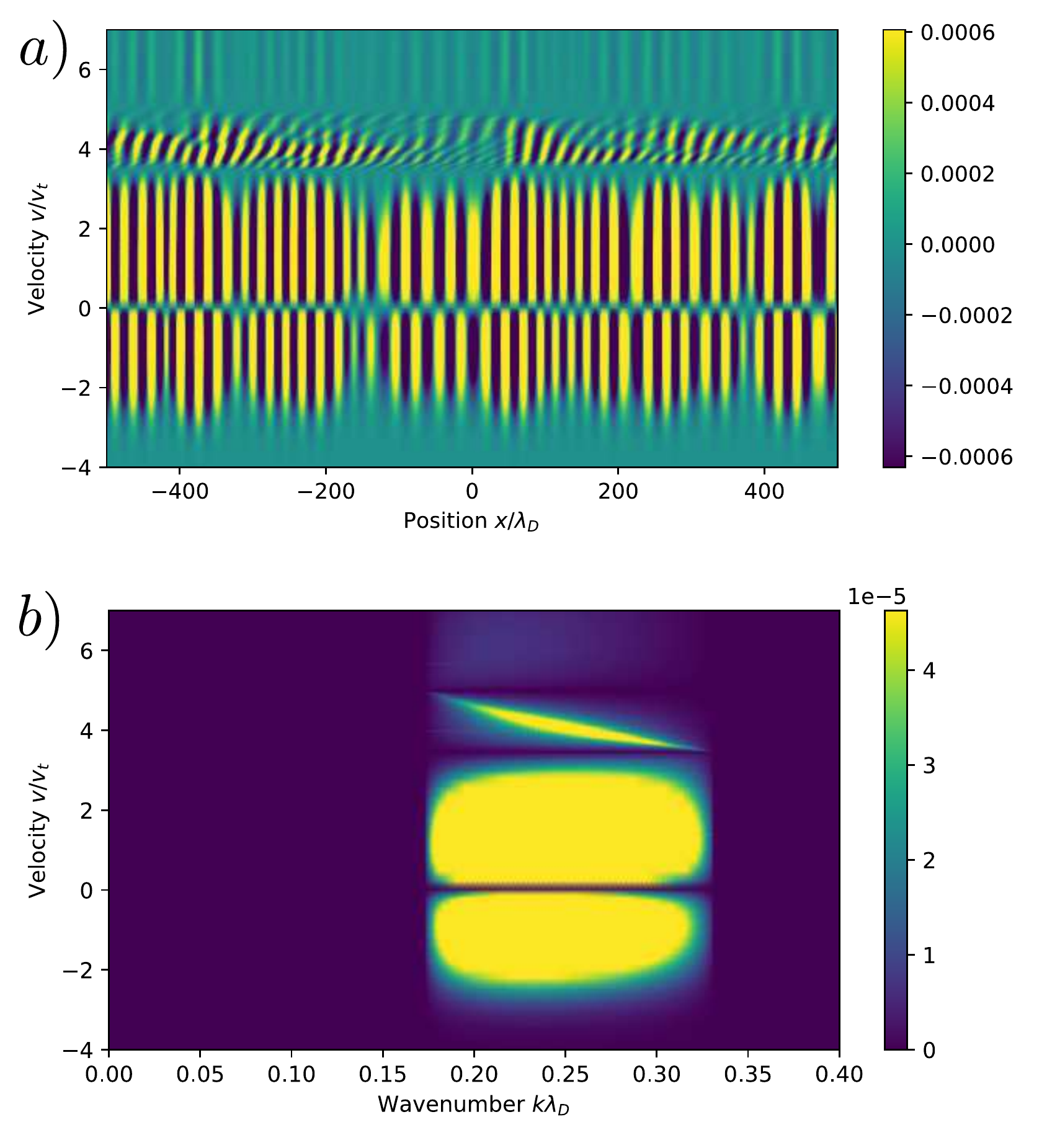}
        \caption{a) The perturbation is shown in phase space representation
            as a subset of the full domain, which has $5000\lambda_D$ length and spans velocities $v\in (-25, 25)$.
            The main thermal body ($v\in(-3,3)$) supports coherent plasma oscillations
            while the modes in the resonant band ($v\in(3,5)$) appear randomly phased.
            The unstable modes are rightward-propagating so the non-resonant distribution of positive velocity $(v\in(0,3))$
            oscillates with greater energy than the negative velocity part.
            Wavepackets in the initial condition are seen as regions of greater amplitude.\\
            b) The Fourier amplitudes as $\log(1 + |f|)$ of the perturbation.
            Spectral energy is peaked along a coherent crescent shape bridging the band of unstable velocities.}
        \label{fig:initial_conditions}
    \end{figure}

    The Vlasov-Poisson simulation is initialized as a sum over the eigenmodes,
    \begin{equation}\label{eq:initial_condition}
            g(x,v) = f_0(v) +
        \sum_n \frac{\alpha_n}{k_n}\text{Re}\Big(\frac{1}{\zeta_n - v}e^{i(k_n x + \theta_n)}\Big)
            \frac{\partial f_0}{\partial v}
    \end{equation}
    where each $\zeta_n$ with $\text{Im}(\zeta_n)>0$ is the solution to $\varepsilon(k_n,\zeta_n)=0$,
    the factor $\theta_n$ is a random phase shift, and $\alpha_n$ is the field amplitude.
    The factor $k_n^{-1}$ sets each initialized mode's field energy to $|E_n|^2 = \alpha_n^2$.
    Figure~\ref{fig:initial_conditions} shows the perturbation on a domain of $L=5000\lambda_D$ initialized
    in this manner.
    The initial condition is chosen in this way to avoid exciting the other damped branches of $\varepsilon(k,\zeta)=0$,
    as the quasilinear theory studies the unstable wave branch.

    \section{Quasilinear theory of the bump-on-tail}\label{sec:quasilinear}
    The quasilinear approximation is the first approximation to velocity-space diffusion in electrostatic turbulence
    and is a reduction from the Vlasov-Poisson theory~\cite{drummond,vedenov_velikhov,diamond,goldston,thorne2017modern}.
    It is mathematically similar to theories of Reynolds-averaging
    and eddy-viscosity in neutral fluid turbulence~\cite{frisch1995turbulence}.
    Consider a spatially-periodic domain of length $L$.
    Let $\langle f\rangle_L\equiv L^{-1}\int_0^L f dx$ be the spatially-averaged distribution and
    $\delta f \equiv f - \langle f\rangle_L$ the fluctuation.
    Equation~\ref{eq:vlasov} is then spatially averaged and the resulting mean equation is subtracted out to obtain an
    equation for the fluctuation.
    The two equations for the mean and the fluctuation are then
    \begin{align}
        \frac{\partial \langle f\rangle_L}{\partial t} &= \frac{\partial}{\partial v}\langle E \delta f\rangle_L\label{eq:background_eq}\\
        \frac{\partial(\delta f)}{\partial t} &+
        v\frac{\partial(\delta f)}{\partial x} = \frac{\partial}{\partial v}\Big(E \langle f\rangle_L
        + E \delta f - \langle E \delta f\rangle_L\Big)\label{eq:quasilinear_pde}.
    \end{align}
    Note that $E=\delta E$ for internal fields.
    The momentum flux of the fluctuation $\delta f$ is composed of two terms: the flux from the mean distribution
    $E\langle f\rangle_L$
    and the fluctuation in the fluctuating flux $E \delta f - \langle E \delta f\rangle_L$.
    The quasilinear closure assumes
    \begin{equation}\label{eq:neglected_term}
         E \delta f - \langle E \delta f\rangle_L \ll E\langle f\rangle_L,
    \end{equation}
    transforming Eq.~\ref{eq:quasilinear_pde} into a quasilinear PDE (one linear in its highest derivatives).
    The quasilinear equations are
    \begin{align}
        \frac{\partial \langle f\rangle_L}{\partial t} &= \frac{\partial}{\partial v}\Big\langle E \delta f \Big\rangle_L\\
        \frac{d(\delta f)}{dt} &= \frac{\partial}{\partial v}\Big(E \langle f\rangle_L \Big)
    \end{align}
    with $\frac{d}{dt} = \partial_t + v\partial_x$ the change along a zero-order trajectory.
    The momentum flux of the background distribution $\langle E \delta  f \rangle_L$ is statistically the un-normalized field-particle
    correlation coefficient~\cite{vankampen} and is a tool increasingly used independent of quasilinear theories
    in the analysis of energy transfer in collisionless plasmas~\cite{howes_klein_li_2017,li_howes_klein_liu_tenbarge_2019,
        klein_howes_tenbarge_valentini_2020,juno_howes_tenbarge_wilson_spitkovsky_caprioli_klein_hakim_2021}.
    The quantity $\langle E \delta f \rangle_L$ also has a basic interpretation as the turbulent momentum flux, in analogy
    to a similar quantity arising in the averaged equation for
    advection of a passive scalar in a fluctuating velocity field~\cite{frisch1995turbulence}.
    These interpretations of $\langle E \delta f \rangle_L$ as a correlation coefficient and as a mean turbulent
    flux are independent of any closure assumptions.

    The quasilinear closure amounts to assuming that the field-particle
    correlation has no variance, or in other words negligible structure inside the averaging box.
    This assumption is often stated as the random phase approximation as there are no wavepackets, or structure, in
    white noise.
    The fluctuating distribution and field are expanded in Fourier series
    to solve the linear equation for $\delta f$ as
    \begin{equation}\label{eq:behavior_of_fluctuating_modes}
        \delta f_n = \frac{iE_n }{\omega(k_n) - k_nv}\frac{\partial \langle f\rangle_L}{\partial v}.
    \end{equation}
    The covariance $\langle E \delta f \rangle_L$ is converted to a spectral sum by Plancheral's theorem,
    \begin{equation}\label{eq:field_particle_plancheral}
        \langle  E \delta f\rangle_L = \frac{1}{L}\int_0^L E \delta f dx = \sum_{n=-\infty}^{\infty} \delta f_n E_n^*.
    \end{equation}
    Substituting Eq.~\ref{eq:behavior_of_fluctuating_modes} into Eq.~\ref{eq:field_particle_plancheral} gives
    the turbulent momentum flux in the quasilinear closure,
    \begin{equation}\label{eq:turbulent_flux}
        \langle E \delta f\rangle_L =
    \Big(\sum_{n=-\infty}^{\infty}\frac{i}{\omega(k_n) - k_n v}|E_n|^2\Big)\frac{\partial \langle f\rangle_L}{\partial v}.
    \end{equation}
    As the left-hand side is the mean momentum flux, and the right-hand side proportional to the mean gradient,
    Eq.~\ref{eq:turbulent_flux} defines a diffusivity as $\langle E \delta f \rangle_L = D(v)\partial_v\langle f\rangle_L$.
    In the sense that the quasilinear closure sets the turbulent flux as proportional to the
    mean gradient this specification of diffusivity is closely
    related to the eddy viscosity theory in physical space dating to Boussinesq~\cite{schmitt2007boussinesq}.
    Now, the simplest equation for the evolution of the spectrum $\mathcal{E}(k)$ is exponential growth
    (assuming no modal interactions and slow modulations),
    where the dispersion relation $\omega(k)=\omega_r(k) + i\omega_i(k)$ is taken as the solution to
    $\varepsilon(\omega, k)=0$.
    From all this one arrives at the electrostatic quasilinear diffusion
    equations~\cite{drummond,vedenov_velikhov,diamond,thorne2017modern,kadomtsev},
    \begin{align}
        \frac{\partial\langle f\rangle_L}{\partial t} &= \frac{\partial}{\partial v}\Big(D(v)\frac{\partial\langle
        f\rangle_L}{\partial v}\Big),
        \label{eq:quasilinear1}\\
        D(v) &=
    \sum_{n=1}^{\infty}\frac{2|\omega_i|}{(\omega_r - k_n v)^2 + \omega_i^2}|E_n|^2,\label{eq:quasilinear2}\\
        \frac{d|E_n|^2}{dt} &= 2\omega_i|E_n|^2,\quad
        \varepsilon(\omega, k_n) = 0,\label{eq:quasilinear3}
    \end{align}
    with the absolute value $|\omega_i|$ following a one-sided time analysis, \textit{i.e.}~causality~\cite{kadomtsev}.
    The equations are intrinsically nonlinear with a self-consistent determination of the dispersion relation
    key to the system's energy and momentum conservation properties~\cite{goldston}.
    The analytic solution of $\varepsilon(\omega, k)=0$ as a complex root can be numerically expensive as $\langle f\rangle_L$ evolves
    from a sum of Maxwellians to a general form.
    The dispersion relation can be solved approximately on the real line using the well-known small growth-rate approximation,
    valid for $\omega_i \ll \omega_r$~\cite{thorne2017modern},
    \begin{align}
        \varepsilon_r(\omega_r, k) &\equiv 1 +
        \frac{1}{k^2}\mathcal{P}\int_{-\infty}^\infty\frac{1}{\zeta_r - v}\frac{\partial\langle f\rangle_L}{\partial v}dv = 0,\\
        \omega_i &= \pi\frac{\partial\langle f\rangle_L}{\partial v}
        \Big(\frac{\partial\varepsilon_r}{\partial\omega_r}\Big|_{\omega=\omega_r}\Big)^{-1}\label{eq:approx_growth_rate}
    \end{align}
    where $\zeta_r = \omega_r/k$ and $\mathcal{P}$ is the Cauchy principal value (P.V.), calculable by Hilbert transform.

    \subsection{The continuous spectrum limit}\label{subsec:continuous_spectrum}
    By defining a spectral density $\mathcal{E}(k) = k_0^{-1}|E(k_n)|^2$ with $k_0 = \frac{2\pi}{L}$ and
    considering the limit as $L\to \infty$, the quasilinear equations can be modeled with a continuous energy spectrum~\cite{thorne2017modern}
    \begin{align}
        \frac{\partial f}{\partial t} &= \frac{\partial}{\partial v}\Big(D(v)\frac{\partial f}{\partial v}\Big),\\
        D(v) &=
        \int_0^\infty \frac{2|\omega_i|}{(\omega_r - kv)^2 + \omega_i^2}\mathcal{E}(k) dk,\\
        \frac{d\mathcal{E}(k)}{dt} &= 2\omega_i\mathcal{E}(k),\quad
        \varepsilon(\omega, k) = 0.
    \end{align}
    The continuous-spectrum approximation is appropriate for large domains and for analysis of the resonant
    and non-resonant contributions to the diffusivity.
    A numerical advantage of using a continuous spectrum is that a high-order polynomial representation may be used
    for the spectral density $\mathcal{E}(k)$.
    The solution of the dielectric function on the spectral points $k_n$ is typically the most expensive part of a
    numerical solution.
    With a high-order representation fewer interpolation nodes are required than the evenly-spaced lattice frequencies
    of the finite-interval problem.
    Fewer evaluations of $\varepsilon(\omega, k)$ speed the simulations considerably.

    \subsubsection{Resonant and non-resonant parts of the diffusivity}\label{subsec:resonant_nonresonant}
    It is common to split the diffusion coefficient into resonant and non-resonant parts
    in the limit of slow modulation $\omega_i\to 0$~\cite{diamond,thorne2017modern,davidson2012methods}.
    Applying the Plemelj relation to the linear response
    \begin{equation}\label{eq:resonant_nonresonant_splitting}
        f_{1,n} \to \mathcal{P}\frac{i}{\omega_r-k_nv}E_n\frac{\partial f_0}{\partial v} +
        \pi \delta(\omega - k_nv)E_n\frac{\partial f_0}{\partial v}
    \end{equation}
    suggests it has two principal terms.
    The first term represents the non-resonant contribution, the response of the main body thermal plasma particles in
    sustaining the wave motion
    (seen in Fig.~\ref{fig:initial_conditions}(a) as the checkerboard pattern for $v\in (-3, 3)$),
    while the second term accounts for resonant energy transfer and is the primary contribution to the
    integration~\cite{kadomtsev, dewar1973oscillation}.
    Having consistently taken the small growth-rate limits, the quasilinear system is usually written as
    \begin{align}
        \frac{\partial f}{\partial t} &=
        \frac{\partial}{\partial v}\Big((D^r(v) + D^{nr}(v))\frac{\partial f}{\partial v}\Big),\\
        \frac{d\mathcal{E}}{dt}&=2\omega_i\mathcal{E},\\
        D^r(v) &= \pi\int_{-\infty}^{\infty}\delta(\omega_r - kv)\mathcal{E}(k)dk,\\
        D^{nr}(v) &= ``\mathcal{P}"\int_{-\infty}^{\infty}\frac{\omega_i}{(\omega_r-kv)^2}\mathcal{E}(k)dk
    \end{align}
    with $``\mathcal{P}"$ the principal value operator.
    Quotes are used because the principal value operation on the denominator, a term quadratic in a nonlinear
    function $\omega_r(k)$, is not necessarily well-defined.
    While it may be tempting to discard the non-resonant diffusivity and evolve only the resonant distribution
    as it is smaller than the resonant diffusivity by approximately $\omega_i/\omega_r$,
    it was well-phrased by Kadomtsev that the population of nonresonant particles is greater by this same factor
    and $D^{nr}$ cannot be neglected~\cite{kadomtsev}.
    The difficulty in a numerical implementation of the resonantly-split quasilinear equations lies in
    approximating the non-resonant diffusivity.
    For example, a typical approximation for $D^{nr}$ is constant in velocity~\cite{diamond,thorne2017modern}.
    For the problem under consideration there is no need to make this approximation.
    The quasilinear problem solved here numerically is the unreduced system of Eqs.~\ref{eq:quasilinear1} and~\ref{eq:quasilinear2}.

    \subsubsection{On analytic prediction of the saturated spectrum}
    Note that as the growth-rate $\omega_i(k_n)\to 0$ the diffusivity corresponding to mode $k_n$ becomes singular,
    \textit{i.e.}~$D(v=\zeta_r(k_n)) = \frac{2}{|\omega_i|}|E_n|^2 \to \infty$.
    As $\omega_i \propto \frac{\partial f_0}{\partial v}$
    the dynamic equation \textit{approaches} a singular diffusion equation in the region of resonant velocities, similar in form to
    \begin{equation}\label{eq:singular_diffusion_equation}
        \frac{\partial f}{\partial t} \to \pi^{-1}\sum_{k_n}
    \partial_v\Big(\varepsilon_r'|E_{k_n}|^2\frac{\partial_vf}{|\partial_vf|_{v=v_{\phi,{k_n}}}}\Big).
    \end{equation}
    Such singular diffusion equations instantaneously flatten the diffusing variable~\cite{giga1999very},
    but the singular problem is approached asymptotically in QL theory.
    The singular behavior of the turbulent diffusivity does not require having taken the small-growth rate
    limit $\omega_i\to 0$ or having considered a resonant/non-resonant splitting of the diffusivity.
    From a numerical standpoint the transition from a regular diffusion equation with smooth diffusivity to a
    near-singular equation with delta-valued diffusivity peaks poses a serious challenge for a numerical
    time-dependent solution of the QL equations.
    Yet this is not necessarily a problem for analysis.

\begin{figure}[htp]
        \includegraphics[width=\columnwidth]{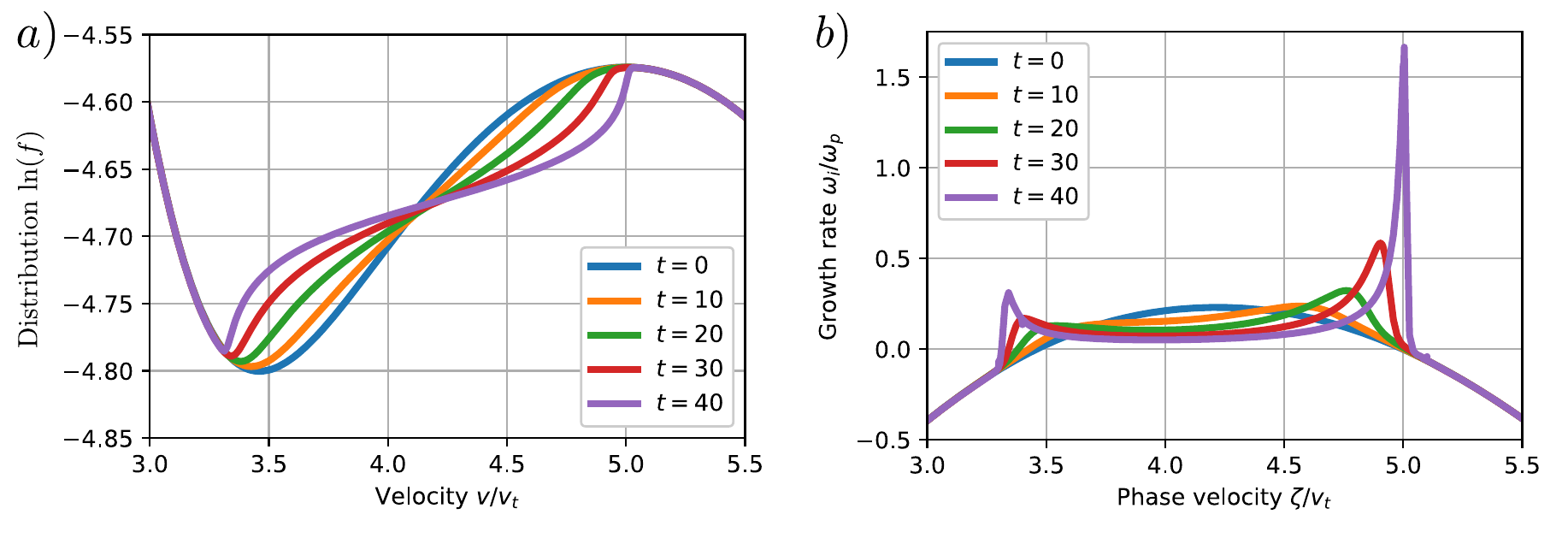}
        \caption{Solution of the quasilinear system using only the resonant diffusion coefficient and assuming
                $\omega=\omega_p$, showing: a) flattening of the distribution function in the region of $\partial_v f > 0$
            and b) modal growth rates.
            Use of only resonant diffusion results in near-singular growth rates at the edges of the diffusing region as
            sharp gradients develop in the distribution.
            This means the growth rate is greatly overestimated, and the saturated spectrum is not correctly
            predicted by Eq.~\ref{eq:predicted_saturation} across the full breadth of wavenumbers, specifically for
            non-initially-resonant phase velocities.}
            \label{fig:resonant_ql_theory}
    \end{figure}

    A theory is sometimes discussed which analytically predicts the
    asymptotic state assuming dispersionless waves $\varepsilon = 1 - \omega_p^2/\omega^2$
    and only resonant interaction, \textit{i.e.} $D^{nr}=0$~\cite{kadomtsev}.
    From this one can show that, with $v_1$ the lower limit of the asymptotically flat region,
    \begin{equation}\label{eq:predicted_saturation}
        |E_k|^2(t_\infty) = \zeta^3\int_{v_1}^{\zeta}f(t_\infty, v') - f(t_0, v'))dv'.
    \end{equation}
    The analysis leading to Eq.~\ref{eq:predicted_saturation} must be cautioned as approximate because the neglect of
    non-resonant diffusivity means that the growth rate at the edges of the diffusing region is significantly
    overestimated, as illustrated in Fig.~\ref{fig:resonant_ql_theory} for the considered bump-on-tail problem.
    Sharp gradients develop in the distribution near the ``notch" and ``hill" of the bump as seen in
    Fig.~\ref{fig:resonant_ql_theory}(a) which would be otherwise filled in by non-resonant diffusion.
    These sharp gradients correspond to near-singular instability growth rates, as shown in Fig.~\ref{fig:resonant_ql_theory}(b).

    \subsection{Conditions for validity of the QL approximation}\label{subsec:validity_conditions}
    A considerable amount of research has been conducted on the conditions under which QL theory should be
    valid~\cite{escande2007can,besse2011validity,elskens2019microscopic},
    aptly summarized in Ref.~[\onlinecite{diamond}] and briefly repeated here.
    The validity conditions center around whether the condition given by Eq.~\ref{eq:neglected_term} is justified, so
    that the fluctuation $\delta f$ should satisfy a quasilinear equation.
    As noted, the condition for a small divergence of the fluctuating flux is unimportance of
    the wavefield's spatial structure within the averaging box.
    Since some distribution of wavepackets from spectral superposition is inevitable, the condition is generally
    rephrased as to whether particles undergo random walks from the wavepacket
    potentials~\cite{diamond}.
    The quantities of importance for this are the electron bounce frequency $\omega_b \approx \sqrt{\rho_c}$ with
    $\rho_c$ the normalized charge density, the wavepacket autocorrelation frequency $\tau_{ac}^{-1} \equiv
    |(\zeta - v_g)\Delta k|$ describing the timescale on which phase waves remain within a wavepacket, and
    the distribution relaxation frequency $\tau_r^{-1} \equiv f_0^{-1}\partial_t f_0$.

    With these quantities, the conditions for validity are that
    i) the bounce frequency be less than the autocorrelation frequency $\omega_b \ll \tau_{ac}^{-1}$,
    and ii) the growth rates lie between the relaxation and autocorrelation frequencies
    $\tau_r^{-1} < \omega_i < \tau_{ac}^{-1}$.
    Since the growth rates $\omega_i\to 0$, this second condition suggests that the QL theory should saturate
    asymptotically with $\tau_r\to \infty$.
    Instability saturation in a finite time violates this condition.

    \section{Numerical simulation methods}\label{sec:discretization_details}
    The 1D1V Vlasov-Poisson system is a two-dimensional hyperbolic-elliptic system, while the QL theory is a 0D1V
    diffusion equation with nonlinear time-dependent diffusivity.
    Thus each model requires its own method of numerical solution.
    This section describes the high-order discontinuous Galerkin methods~\cite{hesthaven2007nodal,shumlak2011advanced,ho2018physics,datta2021electromagnetic}
    developed and applied to each of the two systems.

    \subsection{Discretization of the quasilinear equations}
    The QL diffusion equation is discretized in velocity using the
    local discontinuous Galerkin (LDG) method as high resolution and accuracy
    is needed in the region of resonant velocities when the diffusivity becomes highly peaked.
    A small modification in the LDG method is made to account for the nonlinearity of the system appearing as,
    defining $q\equiv \partial_v f$, the flux $F=D(v)q(v)$ being a product of two interpolated functions.
    The modification from the usual linear LDG method is that the Galerkin matrix elements on the right-hand side of the
    semi-discrete equation are computed by an exact quadrature of the products of the interpolation polynomial basis for
    both $D(v)$ and $q(v)$.

    The grid nodes of high-order DG methods are not evenly spaced, so the Hilbert transform used to compute the
    principal-value integral in the small-growth rate approximation for $\varepsilon(k, \zeta)$ cannot be done using
    FFTs or most other methods discussed in the literature.
    Instead the principal value integral is determined by a quadrature approximation of the velocity Fourier
    coefficients to form the analytic signal, and the P.V.~integral extracted as its imaginary part.
    This is not an optimal approximation but it works sufficiently well on the existing grid provided that the
    distribution is represented to high-enough velocity in the tail regions to be effectively zero.
    For this reason velocity-space is resolved from $v=(-20, 30)$ with one hundred finite elements.
    Each element contains a ninth-order polynomial basis on Lobatto quadrature node interpolants.
    Only twenty elements are used in the tails, with eighty elements spanning $v\in (2,10)$.
    A Fourier-by-quadrature evaluation of the Hilbert transform is not computationally time-limiting as the
    P.V.~integral needs to be evaluated only once per timestep.
    Rather, the nonlinear solution to $\varepsilon(k,\zeta)=0$ for each lattice frequency $k$ is typically the computational
    time-limiting factor.

    With the P.V.~integral determined, the dielectric function $\varepsilon(k, \zeta)$ is
    evaluated as a function of arbitrary phase velocity $\zeta$ via the Vandermonde matrix of the
    quadrature grid's interpolation polynomials.
    After solution with a nonlinear solver to obtain the phase velocity root to $\varepsilon(k, \zeta)=0$
    corresponding to the unstable wave branch,
    the gradient $\varepsilon'(k, \zeta)$ on the interpolated point is found using the local derivative matrix.
    Further computing the interpolated gradient of the distribution function, this determines the approximate linear
    growth rates $\omega_i(k)$ by Eq.~\ref{eq:approx_growth_rate}.
    Since the velocity space is highly resolved the discretization is limited in accuracy by the
    $\mathcal{O}\big(\frac{\omega_i}{\omega_r}\big)$ expansion of $\varepsilon$.

    As the distribution flattens and $\omega_i\to 0$ the diffusivity becomes highly peaked.
    Use of an implicit time integration scheme is necessary to avoid restrictive CFL-limited timestepping and avoid
    dispersive errors, as such oscillations feedback into the growth rate calculation.
    A second-order implicit midpoint method is used to evolve the diffusion equation.
    As the diffusivity nonlinearly depends on the distribution, a nonlinear global solve is avoided by
    taking an approximate explicit half-step in wave energy to the midpoint with a sufficiently small time-step.
    The diffusion equation is then solved by the implicit midpoint method.
    That is, with the semi-discrete QL system as
    \begin{align}
        \frac{df}{dt} &= A(E)f\\
        \frac{dE}{dt} &= 2\omega_i(\partial_v f)E
    \end{align}
    where $A$ represents the discretized diffusion operator and $E=|E_k|^2$ the spectral energy,
    \begin{align}
        E_{n+1/2} &= E_n \exp(h\omega_i(\partial_v f_n)),\\
        A_{n+1/2} &\equiv A(E_{n+1/2}),\\
        f_{n+1} &= (I - \frac{h}{2}A_{n+1/2})^{-1}(I + \frac{h}{2}A_{n+1/2})f_n
    \end{align}
    where $h$ is the time-step.
    The calculations use $h=0.1\omega_{pe}^{-1}$ for the time-advance.

    \subsection{Discretization of the Vlasov-Poisson system}
    The Vlasov-Poisson system is solved using a mixed Fourier/DG spectral method.
    The spatial coordinate is represented on a grid of evenly spaced nodes and the variables $f(x,v,t)$, $E(x)$ are
    projected onto a truncated Fourier series~\cite{boyd}.
    The Fourier coefficients each satisfy
    \begin{align}
        \frac{\partial f_n}{\partial t} &+ ik_nvf_n + \partial_{v}F_n = 0,\label{eq:spectral_vlasov}\\
        F_n(v,t) &= -(E * f)_n,\\
        E_n &= ik_n^{-1}\int_{-\infty}^{\infty}f_ndv,\quad E_0 = 0.
    \end{align}
    An advantage of this formulation is that Gauss's law is solved exactly and computationally is only a constraint
    associated with the system, a familiar approach in incompressible flows.
    The velocity flux $F_n(v,t)$ is computed pseudospectrally using zero-padded FFTs according to Orszag's two-thirds rule
    and is thus de-aliased~\cite{boyd}.
    Velocity space is divided into finite elements and the coupled system of first-order PDEs for $f_n(v, t)$ is
    discretized with the discontinuous Galerkin method using interpolants on Legendre-Gauss-Lobatto (LGL) quadrature
    nodes~\cite{hesthaven2007nodal}.
    As the flux is delocalized in spectral space by the convolution product
    the Lax-Friedrichs flux is used rather than upwinding~\cite{leveque2002finite}.
    The variable-coefficient term $ik_{n}vf_n$ is integrated exactly up to the order of the basis polynomials and is
    thus also alias-free.

    As both spatial and velocity fluxes are alias-free the method has good conservation properties.
    For example, domain-integrated particle number is conserved to machine precision.
    Energy, with no artificial hyperviscosity, is conserved to $\mathcal{O}(10^{-10})$ for the resolution used.
    Spectral blocking, when the turbulent cascade hits the Nyquist frequency of the spatial grid, breaks this energy
    conservation.
    For this reason a hyperviscosity $\nu \nabla_x^4 f$ with $\nu = 1.0\times 10^{-2}$ for $k\lambda_D > 1$
    is used to introduce an artificial cutoff scale at the Debye length.
    This artificial dissipation results in an energy loss $\mathcal{O}(10^{-7})$ by saturation while electric energy
    saturates at $\mathcal{O}(10^{-3})$, meaning the important dynamics are not adversely affected.

    A non-uniform discretization is used, as in the bump-on-tail instability most of the action occurs in the resonant
    band of velocities with $v\in (3,6)$ but velocity space must be truncated at large $v$
    for the above-discussed particle, momentum, and energy conservation.
    Thus only a few elements are used in the tails while most elements are clustered around the main bodies of the two
    Maxwellian distributions.
    In this way sufficient resolution is obtained for fine-scale velocity features without wasted resolution in the tails.
    A semi-implicit time-integrator is used called the AB3CN method~\cite{boyd}.
    In this use of the method, the linear advection term $i k_n v f_n$ is treated implicitly with the second-order
    Crank-Nicholson method while the nonlinear velocity flux is evolved explicitly by the third-order Adams-Bashforth
    multistep method.
    This semi-implicit method is CFL-limited by the spectral velocity flux rather than particle advection, enabling
    arbitrarily large velocity domains.

    In this work $N_v=50$ elements are used to discretize velocity space, each with a ninth-order interpolant polynomial
    basis.
    Velocity space is truncated at $v_{\text{max}}=\pm 25$ thermal velocities in order that total density be
    conserved to machine precision.
    Thirty elements are clustered into the range of approximately resonant velocities from $v\in (3, 8)$ where
    high accuracy is needed to resolve the fine trapping trajectories, with the remainder of elements
    distributed in the main thermal body and tails.
    The physical domain is taken as a periodic box of length of $L=5000\lambda_D$ and is
    divided into $N_x=5000$ evenly-spaced nodes.  

    When the perturbation spectrum is initially peaked at $\max(|E_k|^2) = 10^{-6}$ the instability saturates at
    $t\omega_p\approx 130$.
    The simulation is run to $t_{\text{max}}=170$ with a $\Delta t = 2.5\times 10^{-3}$,
    requiring about fifteen minutes on a single RTX 3090 GPU\@.
    For a given time-advance the numerical results are well converged, such that discretization errors are below machine
    precision as verified by varying the resolution.
    However some phase error is introduced by the implicit time integration of spectral velocity advection as the very
    smallest scales are slowed to the chosen time-advance~\cite{boyd}.
    As these smallest modes are damped by the hyperviscosity no significant difference is seen in domain averages
    such as $\langle E\delta f\rangle_L$ when compared to explicitly time integrated simulations.
    This error does mean that care should be taken with the choice of time-step as small scales are slowed in addition
    to large velocities.


    \section{Analysis of simulation results}\label{sec:sim_results}
    Here the Vlasov-Poisson and quasilinear simulations are analyzed, and the discrepancy between the two simulations
    explored.
    Both results use the initial background distribution specified in Eq.~\ref{eq:unstable_distribution} and suppose
    a domain length of $L=5000\lambda_D$.
    Section~\ref{subsec:choice_of_spectrum} makes some comments on the choice of initial condition.
    A comparison of the turbulent flux is made in Section~\ref{subsec:diffusivity_evolution}, along with
    a discussion on the fluctuations in the field-particle correlation $\langle \delta f E\rangle_L$ responsible
    for deviation from quasilinear diffusion.
    Section~\ref{subsec:eddy_analysis} analyzes the role of phase space structures in the solution of the Vlasov
    equation followed by an analysis of the validity conditions as discussed in Section~\ref{subsec:validity_conditions}.

%
%
%

    \begin{figure}[htp]
%
        \includegraphics[width=\columnwidth]{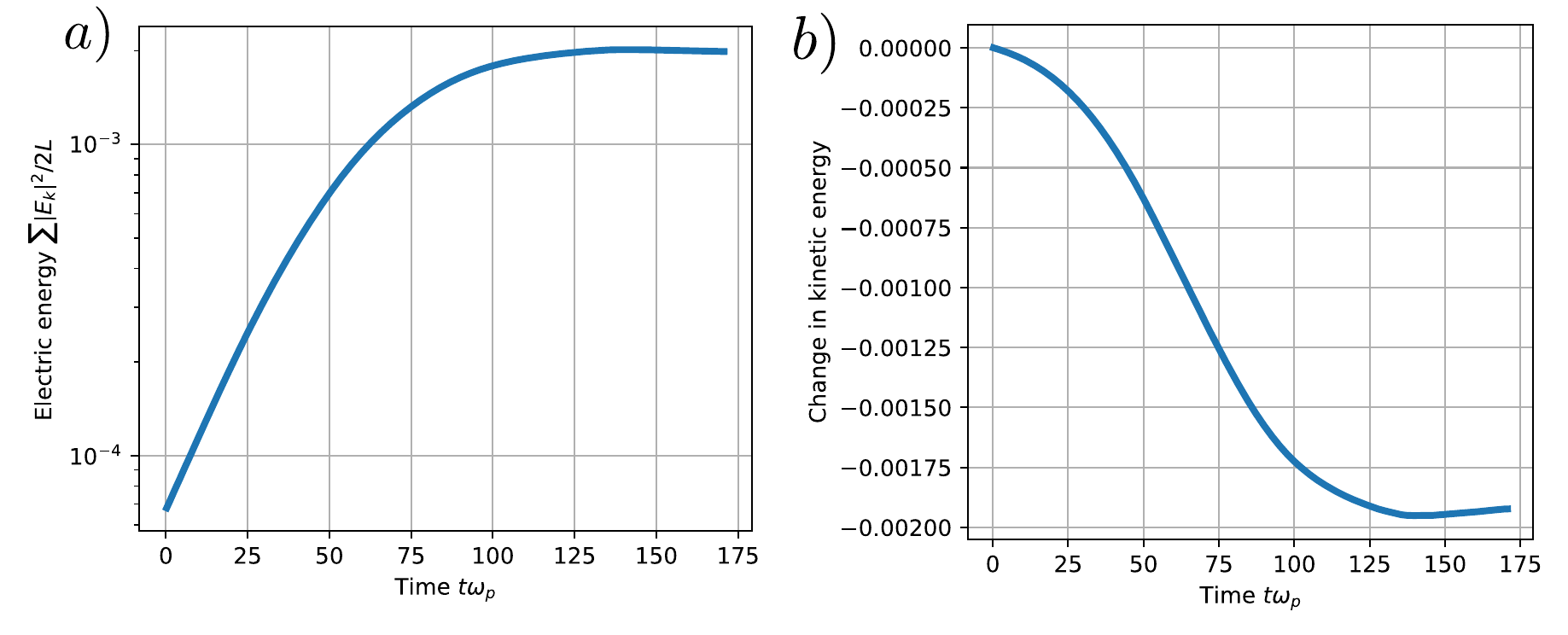}
         \caption{Domain-integrated energy traces for the Vlasov-Poisson simulation of: a) electric field energy and
         b) change in kinetic energy. The bump relaxes as electrons release free kinetic energy.
            A typical feature of nonlinear saturation is an overshoot of equilibrium and slow nonlinear evolution,
                seen here as an increase in kinetic energy past saturation at $t\omega_p \approx 125$.
            An artificial hyperviscosity is applied for wavenumbers $k\lambda_D > 1$, but total energy per unit
             length is conserved to $\mathcal{O}(10^{-7})$.}\label{fig:energy}
    \end{figure}

    \subsection{On the choice of initial energy spectrum}\label{subsec:choice_of_spectrum}
    The dynamic behavior of the quasilinear system as an initial value problem depends on the choice of initial
    condition, because the spatially-averaged distribution $\langle f\rangle_L$ and also the initial energy spectrum $|E_k|^2$
    are free to be specified.
    All modes must be energized as change takes place only through linear growth or damping.
    A variety of choices exist for the initial distribution of spectral field energy $|E_k|^2$.
    A possible choice is a constant energy in all modes $E_k$ with random phases, \textit{e.g.}~a white noise.
    Yet one cannot put energy into damped ``eigenmodes" of the Vlasov-Poisson system without exciting the other
    branches of the kinetic dispersion relation, as Landau-damped modes are not true eigenfunctions~\cite{case1959plasma}.
    If initialized, damped kinetic modes can couple into the dynamics of growing modes.
    For example, Fig.~\ref{fig:energy} shows the temporal change of domain-integrated electric and kinetic energies in
    the development of the instability.
    If perturbations other than eigenmodes are initialized then the domain-integrated energy will oscillate as
    it grows in time.
    The Vlasov-Poisson simulation is perturbed only by unstable modes for better comparison to quasilinear theory.

    Initialization of constant energy across all modes results in a cusped diffusivity profile,
    nearly discontinuous at the edges of the resonant velocities, because some wavenumbers have near-zero growth rates.
    Because of this cusp effect the energy spectrum is initialized in both simulations to match a scaled
    version of the positive growth rates with the maximum $\alpha_n^2=10^{-6}$ in Eq.~\ref{eq:initial_condition}.
    In the quasilinear system the damped part of the spectrum is initialized to the minimum positive growth rate
    in order that the initial spectrum is continuous.

    \begin{figure}[htp]
        \includegraphics[width=\columnwidth]{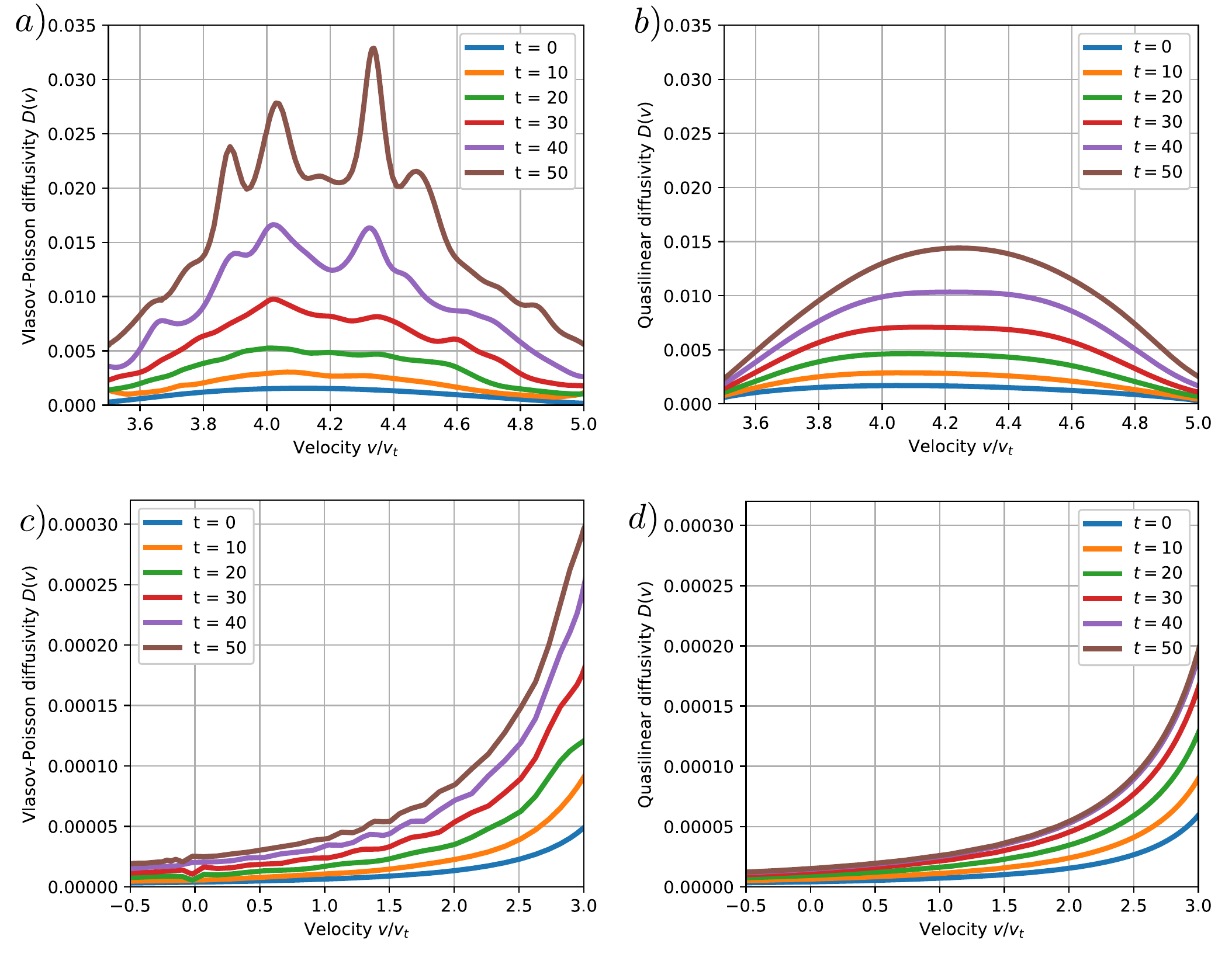}
        \caption{Early-time evolution of diffusivity $D(v)$ in: (a-b) the region of resonant velocities;
        and (c-d) for non-resonant velocities in the main thermal body.
        The Vlasov-Poisson diffusivity in (a) estimated with Eq.~\ref{eq:estimated_diffusivity}.
            The diffusivities agree well until around $t\omega_p=40$, where they diverge significantly.
            A key difference is the width of the diffusing region, as the Vlasov simulation diffusivity at the
            resonant edges ($v=3.5$ and $v=5$) is twice as large as the quasilinear prediction.
            Note that wave energies, plotted in Fig.~\ref{fig:energy}, are not yet at nonlinear levels.
            By $t\omega_p=50$ non-resonant diffusivity of the Vlasov-Poisson simulation in (c) is roughly $30\%$ greater
            than the diffusivity in (d) predicted by quasilinear theory for the $L=5000\lambda_D$ interval.}\label{fig:early_resonant_diffusivity}
    \end{figure}

    \subsection{Comparisons of turbulent diffusivity and momentum flux}\label{subsec:diffusivity_evolution}
    Primary quantities of interest in phase space turbulence are the velocity-space diffusivity and the turbulent
    momentum
    flux $\langle E \delta f \rangle_L$, in particular its profile and magnitude, in the regions of resonant and
    non-resonant velocities.
    Based on Eq.~\ref{eq:turbulent_flux}, a turbulent diffusivity can be estimated from the Vlasov-Poisson simulation
    by decomposing the distribution into a component that is averaged over physical space $\langle f\rangle_L$
    and a component fluctuating about this average, $\delta f = f - \langle f\rangle_L$.
    Then diffusivity is estimated as
    \begin{equation}\label{eq:estimated_diffusivity}
        D_{\text{estimated}}(v) \equiv \frac{\langle E \delta f \rangle_L}{\partial_v \langle f\rangle_L}
    \end{equation}
    as the turbulent momentum flux normalized to the gradient of the background distribution.

\begin{figure}[htp]
        \includegraphics[width=\columnwidth]{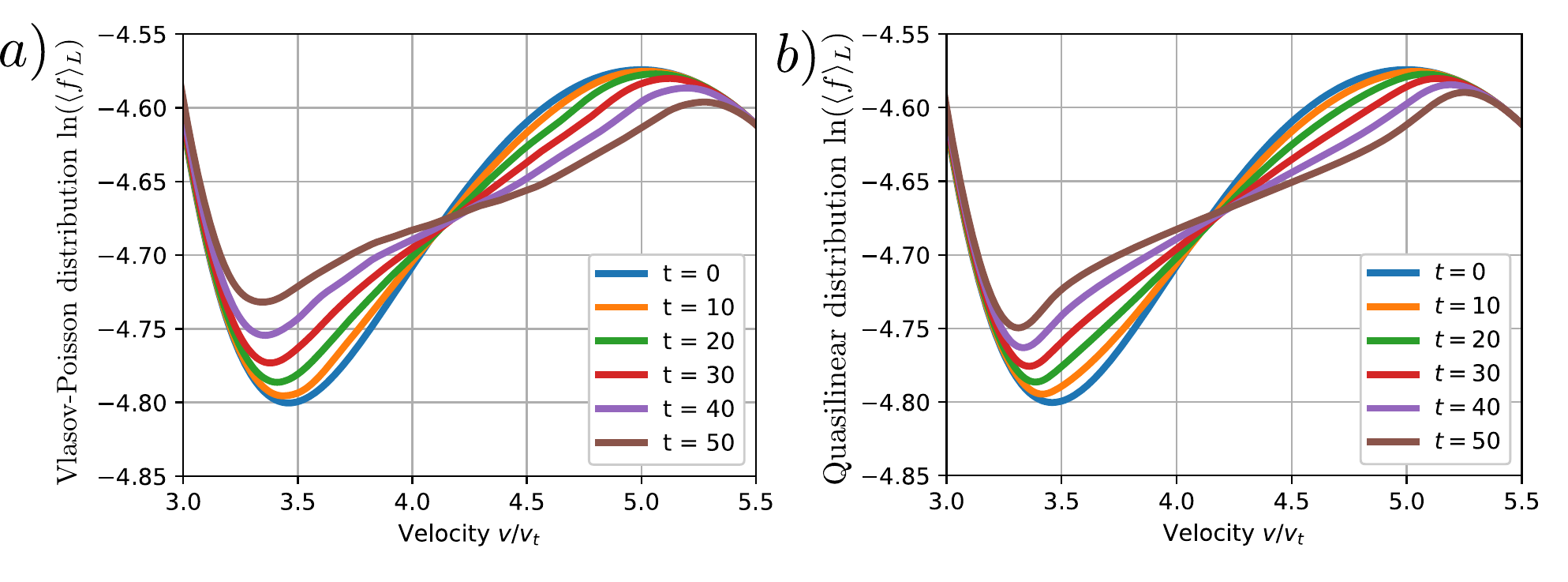}
        \caption{Diffusion of the distribution function in the tail region.
        Up to this time the quasilinear solution agrees closely with the Vlasov solution, though by $t\omega_p=50$ the
        Vlasov-Poisson simulation has a noticeably smoother profile as the diffusivity has grown significantly greater
        than the quasilinear value.
        By $t\omega_p=90$ the Vlasov simulation has completely filled in the low-velocity notch, while
        the quasilinear simulation maintains a steep notch area.}\label{fig:early_flattening}
    \end{figure}

    Figure~\ref{fig:early_resonant_diffusivity} shows the Vlasov-Poisson
    and quasilinear velocity-space diffusivities in the resonant (a-b) and non-resonant regions (c-d),
    demonstrating that quasilinear theory correctly predicts the shape
    and early development of the profile of turbulent diffusivity, but later presents an underestimate.
    By $t\omega_p=50$ the velocity-space diffusivity is underestimated by between 30--100\%.
    Regarding the non-resonant diffusivity, Fig.~\ref{fig:early_resonant_diffusivity}(c-d) shows that this value
    varies by an order of magnitude across the positive velocities of the main thermal body, becoming smaller
    for negative velocities.
    The variation of diffusivity with velocity is important for the change in $\langle f\rangle_L$ in the tail region,
    with the two simulations compared in Fig.~\ref{fig:early_flattening}.
    The main thermal body is non-resonantly heated primarily in positive velocities.
    The ``notch" or depression between the main thermal body and the bump in $\langle f\rangle_L$ is in this way filled
    by diffusion of the distribution from near-resonant, but not fully resonant, velocities.

\begin{figure}[hbp]
        \includegraphics[width=\columnwidth]{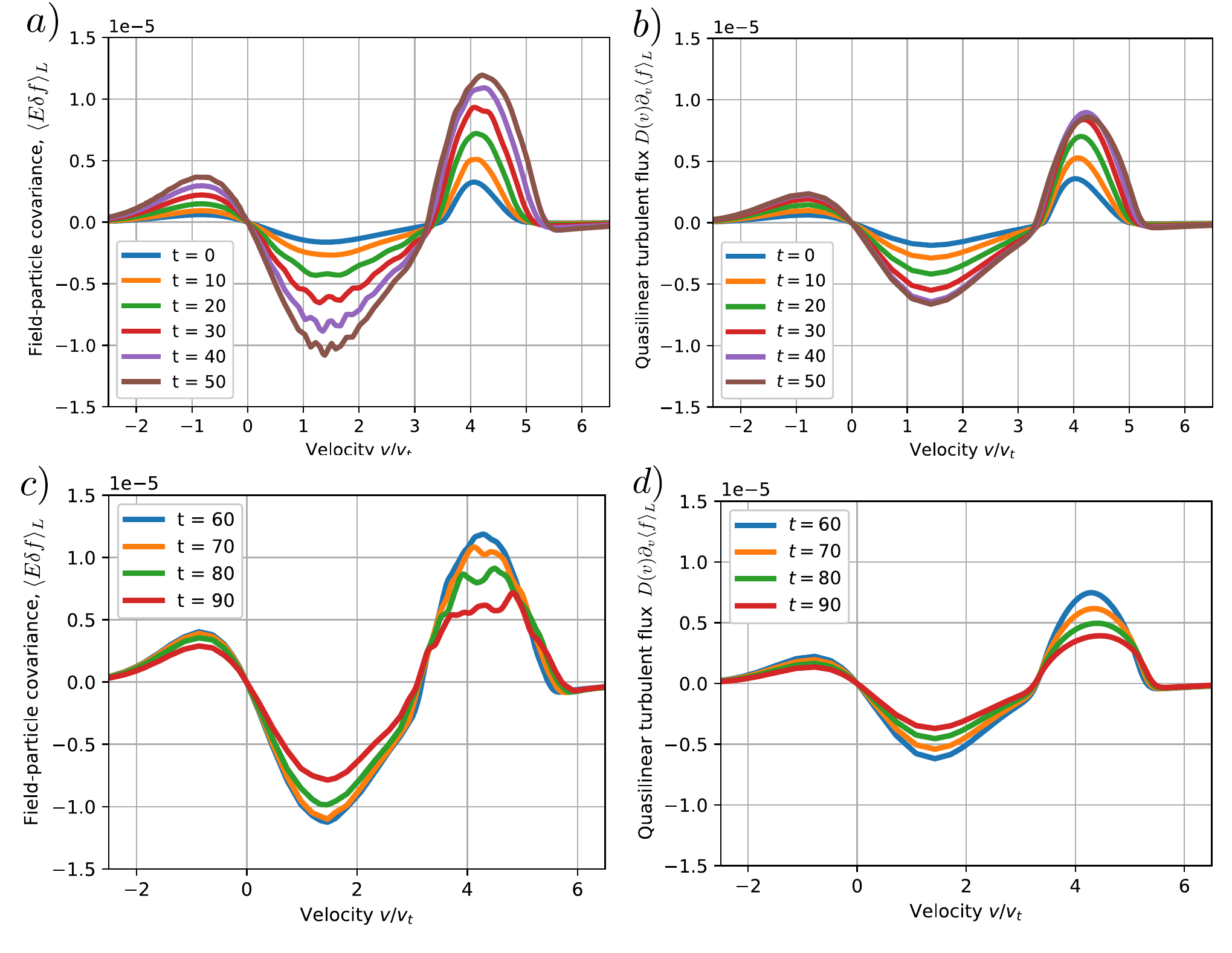}
            \caption{Time-evolution of the field-particle correlation, or equivalently turbulent
        momentum flux of the averaged distribution function, for (a, c) the Vlasov-Poisson solution
            and (b, d) the quasilinear solution.
        The turbulent flux of both systems peaks around $t\omega_p=50$ for this initial condition, yet the observed
            flux of the Vlasov simulation is significantly greater than the quasilinear value
            for all times beyond $t\omega_p = 30$.}\label{fig:turbulent_flux}
    \end{figure}

    For later times in the Vlasov simulation fluctuations of
    $\langle f\rangle_L$ make the diffusivity estimate unreliable as its gradient oscillates.
    This happens to occur as the two models begin to diverge and the diffusive closure breaks down.
    A smoother quantity to compare is the covariance, or turbulent flux,
    $\langle E \delta f \rangle_L$.
    Figure~\ref{fig:turbulent_flux} compares the quasilinear turbulent flux to that of the
    Vlasov-Poisson simulation.
    This figure shows that the flux in the region of non-resonant velocities close to the
    resonant region around $v\approx [3,6]$ is underestimated by quasilinear theory.
    This enhanced turbulent flux flattens the extreme regions of the unstable distribution, meaning the
    ``notch" and ``bump" of the evolving tail.
    These two regions are flattened asymptotically in the quasilinear model.
    The effect of this enhanced diffusion is to quench the instability in a finite time,
    in contrast to quasilinear theory's asymptotic approach to quenching.
    This is seen as the saturation of electric energy in Fig.~\ref{fig:energy}(a).

\begin{figure}[htp]
        \includegraphics[width=0.75\columnwidth]{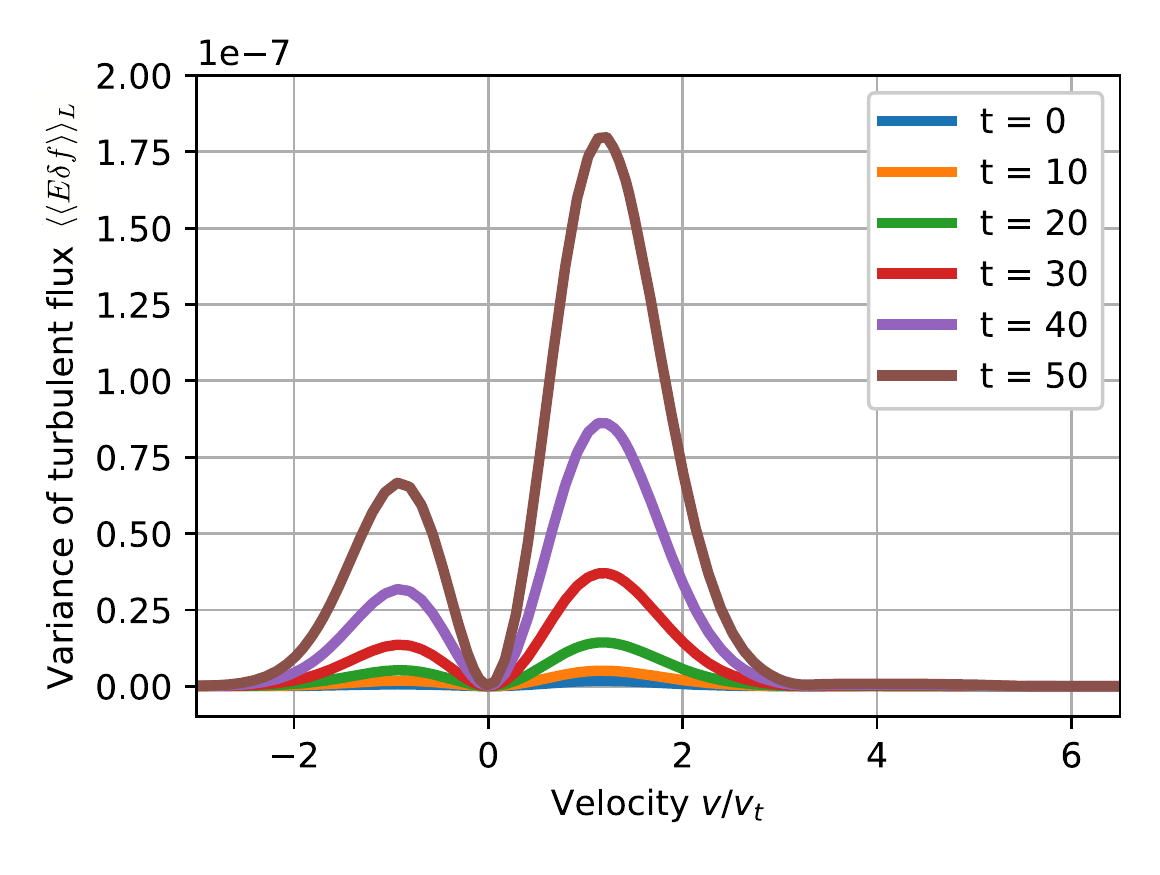}
        \caption{Evolution of the variance of the turbulent flux, computed as the mean-square in the fluctuating flux
        $E \delta f - \langle E \delta f\rangle_L$, showing significant values in non-resonant velocities only.
        The variance of the turbulent momentum flux for non-resonant particles by two orders of magnitude over resonant
            ones supports the mechanism for the disagreement with the quasilinear theory as due to spatial structure
        in the electric field, \textit{i.e.} beat waves.}\label{fig:variance_of_field_particle_correlation}
    \end{figure}

    Another tool to quantify the quasilinear closure is the mean-square of the fluctuation in the
    turbulent flux, \textit{i.e.}~the variance $\langle\langle E \delta f  \rangle\rangle_L$ plotted in
    Fig.~\ref{fig:variance_of_field_particle_correlation}.
    Recall the closure assumption Eq.~\ref{eq:neglected_term} comparing a 0D1V-dimensional quantity to
    a 1D1V-dimensional one.
    This comparison may be valid in some areas and invalid in others.
    The variance quantifies the fluctuations in the turbulent
    flux and points out which velocities contribute most to breaking the assumed closure.
    Most variance occurs in the energy transfer corresponding to the non-resonant distribution.
    This supports the evidence from Fig.~\ref{fig:turbulent_flux} that a modification of the turbulent diffusivity
    due to spatial structure, meaning wavepacket potentials, is primarily
    responsible for the deviation of the turbulence evolution from the quasilinear prediction.

    \subsection{Analysis of phase space structures in the Vlasov-Poisson simulation}\label{subsec:eddy_analysis}
    Section~\ref{sec:quasilinear} reviewed quasilinear theory and highlighted parallels to models of eddy
    viscosity in the turbulence of an incompressible fluid.
    In such eddy viscosity models a passive scalar is mixed by its partial participation in an ensemble of fluid eddies,
    resulting in a diffusive character to the scalar's averaged profile.
    By observing the phase flow within the averaging box a similar
    structure of eddies is observed due to phase fluid resonantly interacting with phase waves.

\begin{figure}[htp]
    \includegraphics[width=\columnwidth]{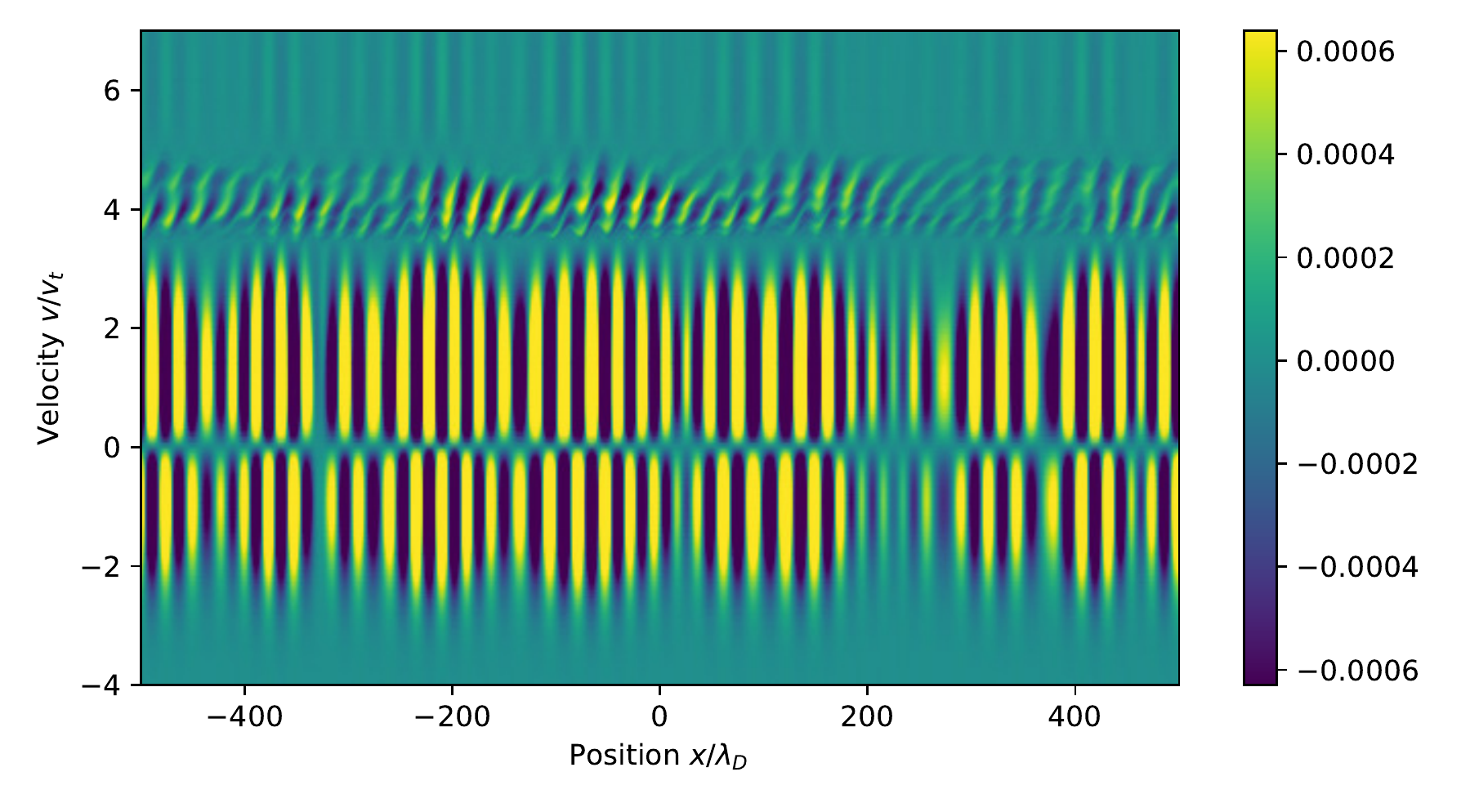}
    \caption{The fluctuation in the distribution function, $\delta f \equiv f - \langle f\rangle_L$,
            of a Vlasov simulation is plotted during the quasilinear phase of evolution at $t\omega_p = 10$.
            A short domain ($L=1000\lambda_D$) is shown in order that the whole flow can be observable at once.
            A clear distinction can be made between resonant and non-resonant parts of the fluctuation.
            Non-resonant velocities undergo coherent oscillations (seen as the striped pattern) while phase fluid
            at resonant velocities ($v\in (3,5)$) is accelerated by randomly-phased waves.
        These structures are consistent with quasilinear evolution.
        }\label{fig:results0}
    \end{figure}

    Section~\ref{subsec:diffusivity_evolution} shows that the Vlasov simulation evolves in two phases, at first in
    agreement with quasilinear diffusion and later diverging as wave energy increases.
    Figure~\ref{fig:results0} shows the development of the fluctuation $\delta f$
    at an early time corresponding to quasilinear evolution for a domain of $L=1000\lambda_D$.
    Phase fluid mixes without the formation of closed eddies.
    A shorter domain was chosen here to present a complete picture which can be reasonably
    viewed as a whole.
    The simulation in Fig.~\ref{fig:results0} uses $5000$ spatial nodes and does not use spatial hyperviscosity.
    Note that the checkerboard pattern of the plasma oscillations is similar to the eigenmodes
    discussed in Section~\ref{sec:linear_theory} and visualized in Fig.~\ref{fig:initial_conditions}.

    The Vlasov simulation begins to disagree with quasilinear theory as phase space eddies form.
    These eddies are also referred to as clumps or granulations and are a central ingredient of the theory
    of phase space structures~\cite{dupree1970theory,
        dupree1972theory,boutros1981theory}.
    These structures can be roughly divided into coherent (eigenmode-like) and incoherent (eddy-like) parts.
    This section discusses the dynamics of the eddy-like part of the fluctuation.
    An extensive textbook discussion of granulations and their role in phase space turbulence
    can be found in Chapter 8 of Ref.~[\onlinecite{diamond}].
    Of course the analogy to eddy-viscosity models is only partial since phase space eddies account only for resonant
    diffusion;
    it is well-known from theory  
    that coherent oscillations of the main thermal body are responsible for energy transfer at
    non-resonant velocities~\cite{dewar1973oscillation,cary1981ponderomotive}.

    \subsubsection{Eddy analysis of the direct Vlasov simulation}
\begin{figure}[hbp]
        \includegraphics[width=\columnwidth]{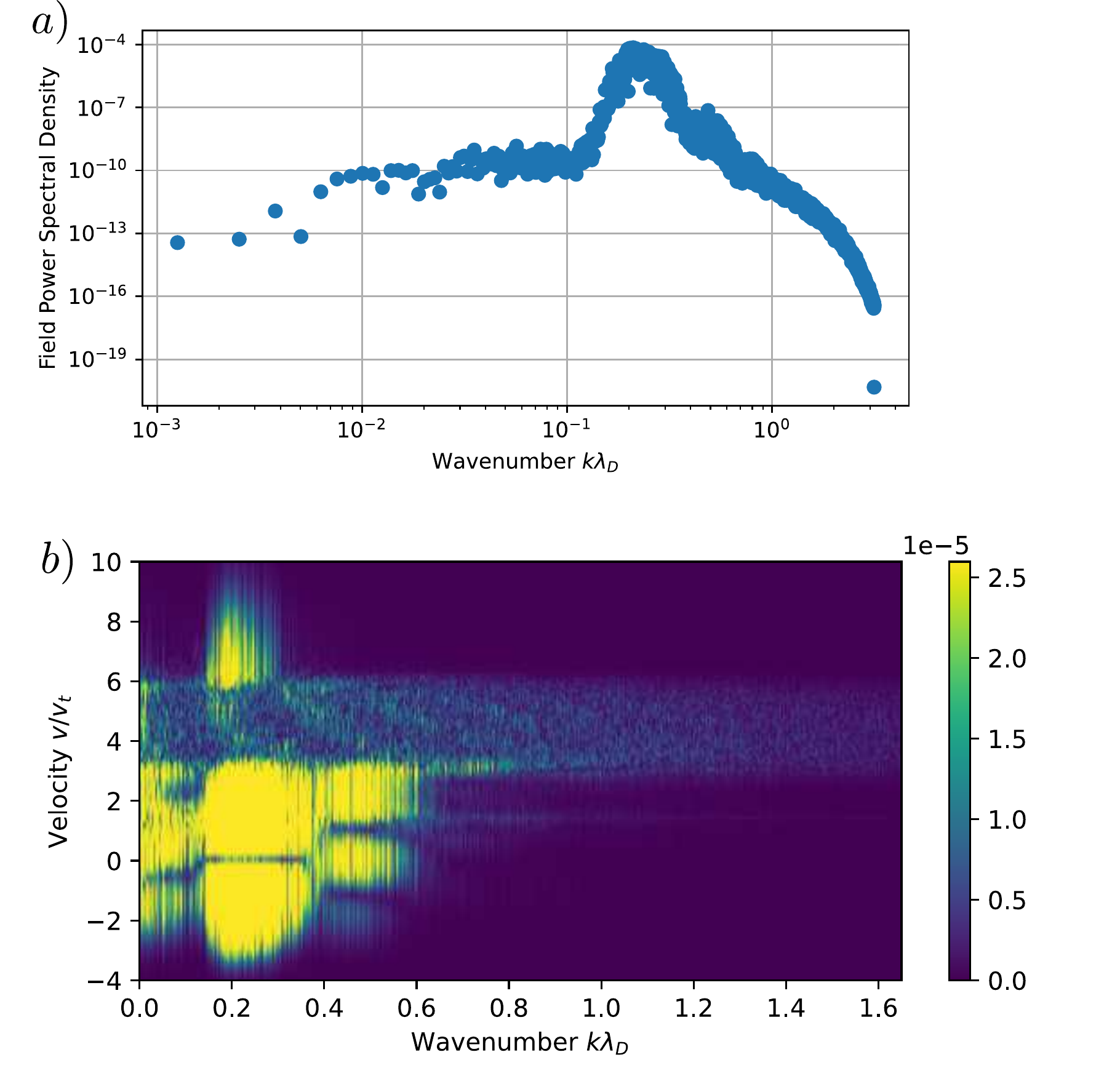}
        \caption{a) The field spectrum $|E_k|^2$ of the Vlasov simulation averaged from $t\omega_p=140$--$170$,
        consisting of two peaks at $k\lambda_D\approx 0.25$ and $0.5$, as well
            an inertial range-like power law for $k\lambda_D\gtrsim 0.6$.
            The primary peak is due to the saturated instability and the secondary peak to its first harmonic.
            The spectral knee and dissipation around $k\lambda_D=2$ is due to the artificial hyperviscosity.\\
        b) Fourier-decomposition of the distribution function, shown in logarithmic view as $\log(1+|f|)$,
            demonstrating that the high-$k$ power law
                is due to eddy turbulence of resonant electrons,
                while the energy-containing oscillations are present in the population of
                non-resonant electrons.}\label{fig:dns_spectra}
    \end{figure}

    One might expect each linear mode to evolve independently until terms nonlinear in the wave amplitude
    dominate the dynamics.
    Yet as seen in the Fourier spectrum of the fluctuation in Fig.~\ref{fig:dns_spectra}(b),
    instability development leads to a broadening of the initialized coherent crescent
    shape in the distribution spectrum $f(k,v)$ and the gradual development of a power law in these Fourier
    coefficients around the resonant band of velocities.
    Initial broadening of the Fourier spectrum in the linear phase of the instability, representing evolution of
    fine structures in the band of resonant velocities from reaction on the mean momentum flux, is the essence of
    the quasilinear effect.
    However the power law seen in Fig.~\ref{fig:dns_spectra}(a) represents turbulent features not captured by the
    quasilinear approximation.

    An analysis of phase space eddies reveals how quasilinear closure breaks down
    as this resonant-velocity power law from turbulent eddy mixing develops.
    Eddies consist of electron phase fluid oscillating in a local potential well.
    The typical electron bounce frequency at saturation, measured as the rms charge density fluctuation,
    is $\omega_b\approx 0.1\omega_p$, extending to bounce frequencies of $\omega_b\approx 0.2$--$0.4\omega_p$ in the largest
    amplitude wavepackets.
    This gives a range of eddy turn-over times from $\tau_b\omega_p\approx 15$--$60$.
    Resonant particles do not experience a full period of bounce motion, as phase fluid transits each potential well
    in a time
    \begin{equation}\label{eq:potential_transit_time}
        \tau_{\text{transit}} \equiv \frac{\lambda}{v-v_g}
    \end{equation}
    with $v$ the local phase fluid velocity, $v_g$ the wavepacket group velocity, and $\lambda$ the width of a wavepacket
    potential trough.
    In the simulation considered the width of a wavepacket potential trough is roughly $\lambda \approx 10\lambda_D$,
    the group velocity of wavepackets roughly one thermal velocity, and the velocities of resonant particles
    $v \approx (3, 6)v_t$ (here estimated as greater than the original $v\in (3, 5)v_t$ as the resonant region expands
    with time).
    This gives transit times of $\tau_{\text{transit}}\omega_{p}=2$--$5$.
    Thus groups of resonant particles in the mixing layer participate in between $3$--$30\%$ of a typical phase space
    eddy motion per trough.
    However, since the potential profile is similar in each wavepacket trough, and the clump enters the trough
    approximately as it left, a rotating clump can turn over in the process of passing through multiple potential
    troughs.
    In this way a clump of phase fluid is progessively distorted by its
    passage through the potential profile of a wavepacket and trajectories randomized by the process of phase mixing
    in a nonlinear potential.

\begin{figure}[htp]
%
        \includegraphics[width=0.75\columnwidth]{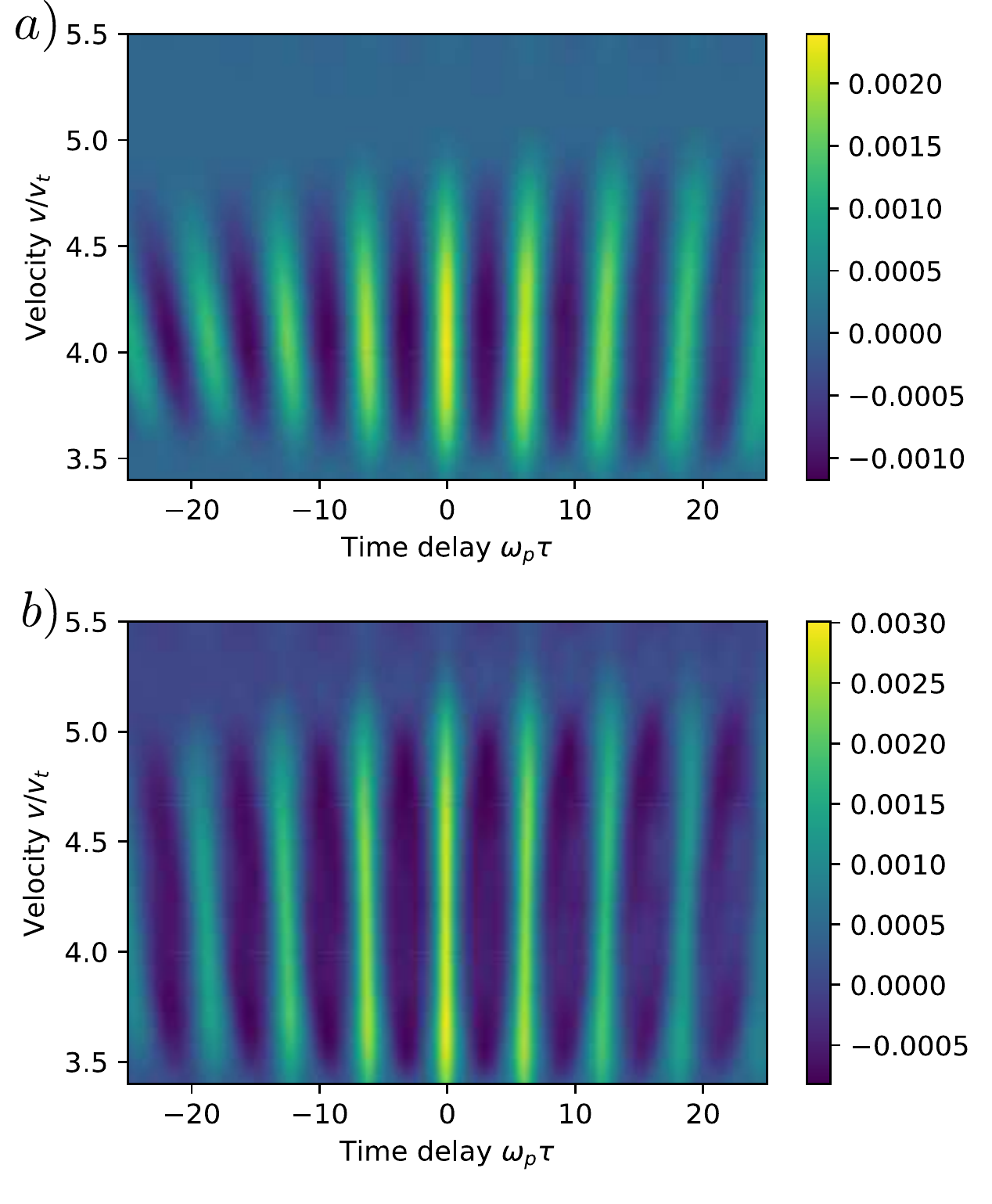}
        \caption{Autocorrelation $\langle \delta f(x,v,t)\delta f(x-v_g\tau,v,t-\tau)\rangle_L$ of
            the fluctuating distribution function in the Vlasov-Poisson simulation as a function
            of time-delay $\tau$ plotted for: (a) the quasilinear phase of instability
            evolution, and (b) the development of anomalously high diffusivity.
            A sinusoidal signal is indicative of an underlying oscillatory pattern,
                while a positively-peaked signal indicates a repeating pattern.
            As the instability evolves this pattern grows increasingly peaked.
                By saturation there is only recurrence with no oscillatory signal in resonant velocities.
            The oscillatory signal occurs during times in agreement with quasilinear theory,
                while recurrence coincides with the increase of turbulent flux beyond the
                quasilinear value.
            The resonant fluctuation recurs at the plasma frequency, decorrelating into the past and future over
            about six plasma periods.
            The decay in autocorrelation peaks corresponds to the approximate coherence time of phase space
                clumps.}\label{fig:autocorrelations}
    \end{figure}

\begin{figure}[htp]
        \includegraphics[width=0.9\columnwidth]{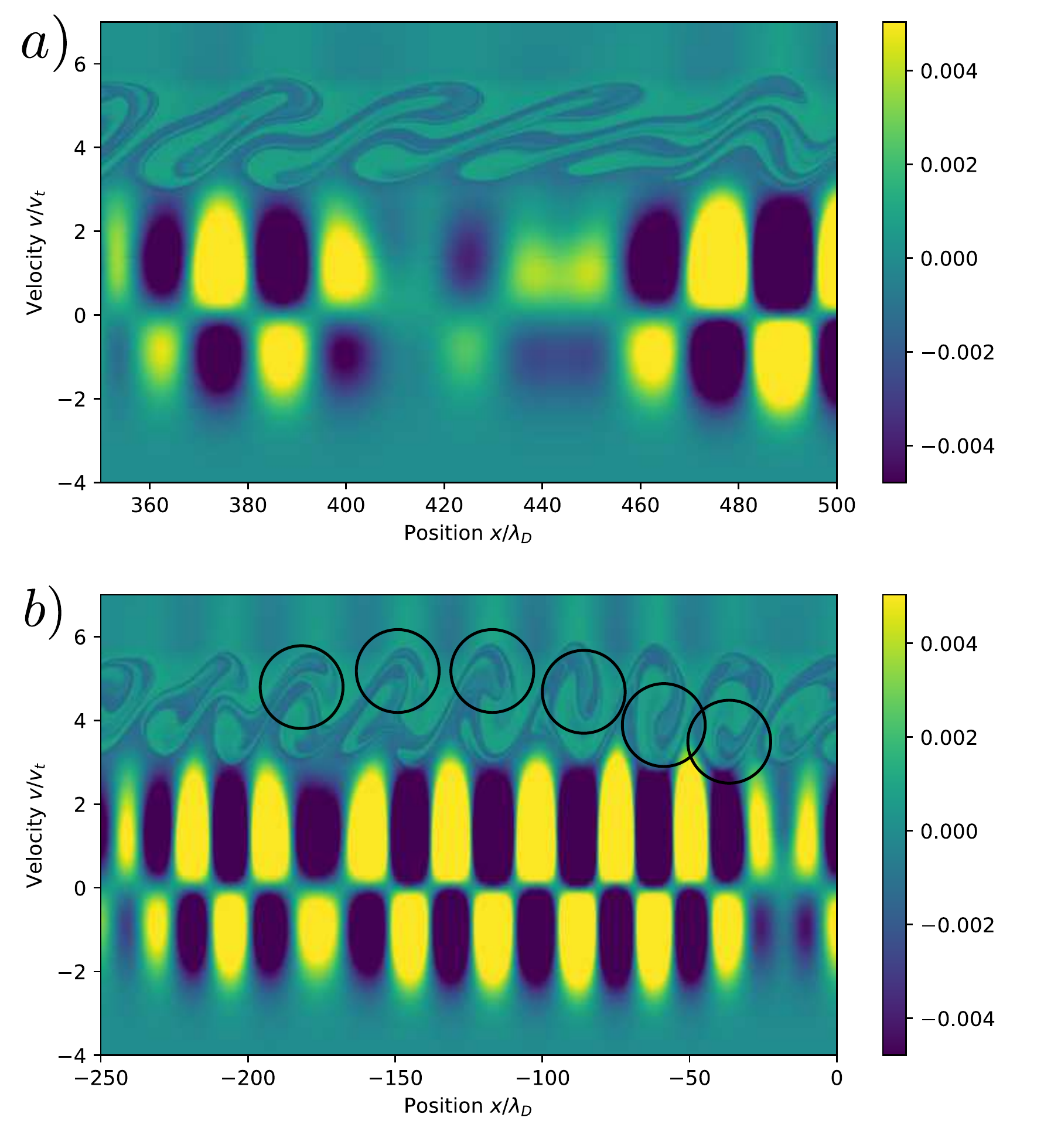}
        \caption{Detailed phase-space views of the fluctuation $\delta f \equiv f-\langle f\rangle_L$ in the Vlasov-Poisson
        simulation:
        a) outside a wavepacket; and b) within a wavepacket.
            Phase space is shown at $t\omega_p=120$ for the $1000\lambda_D$ domain.
            Null-points between wavepackets can be determined from the phase-space view as the regions
            where the coherent plasma oscillation sustained by non-resonant electrons goes to zero amplitude.
            The mixing mechanism depends on the local potential structure.
            Outside wavepackets filaments mainly free-stream while within wavepackets eddies mix.
            The long decorrelation time of a clump is illustrated in (b) by a circle tracking a recurrent clump,
            demonstrating that an eddy can complete a turn as it propagates from trough-to-trough, even though the
            rotating clump is experiencing different potential amplitudes in each trough.}\label{fig:results_zoom}
    \end{figure}

    A recurrence effect allows this progressive eddy turnover to be visualized in a single-time snapshot of the phase
    flow.
    To understand the recurrence effect, consider the fluctuation's autocorrelation function
    $\langle \delta f(x,v,t)\delta f(x-v_g\tau,v,t-\tau)\rangle_L$~\cite{diamond},
    plotted in Fig.~\ref{fig:autocorrelations} for resonant velocities during early quasilinear evolution and a later
    time when evolution deviates from quasilinear diffusion.
    During the quasilinear phase the autocorrelation signature is sinusoidal but during later evolution
    the signature changes to that of a structure repeating at the plasma frequency.
    The repeating signal corresponds to a temporal similarity in the phase flow.
    That is, one can approximately trace the past and future behavior of a phase space clump in a given potential well
    by looking at the neighboring potential wells because clumps transit between wells in approximately one plasma period.
    Figure~\ref{fig:results_zoom}b shows this effect in detail, tracing the approximate future state of a clump as it
    transits a wavepacket with a series of small circles.
    In the way visualized, resonant particle mixing becomes governed by progressive eddy turnover through the distinct
    potential wells of a wavepacket.
    The eddies visualized are not trapped in any particular potential well.

    The phase fluid patterns have a long coherence time because the degree of phase mixing in a sinusoidal
    potential depends on the time spent participating in the eddy motion.
    Particle trajectories in a given clump slowly decorrelate, with a decorrelation time given by the decay of peaks
    in Fig.~\ref{fig:autocorrelations}.
    The change from a sinusoidal signal to that of a repeating structure in the autocorrelation function at resonant
    velocities is concurrent with the time of disagreement of the turbulent flux from the quasilinear prediction.
    This is the time for which the particle bounce frequency (or equivalently, phase space eddy turnover time)
    becomes smaller than but comparable to the wavepacket autocorrelation and resonant particle
    transit times.

    \subsubsection{Wavepacket autocorrelation and quasilinear applicability}
    Besides its oscillation frequency a wavepacket is characterized by its autocorrelation time $\tau_{\text{ac}}$, or
    the lifetime of the potential profile in the wavepacket, and the lifetime of the wavepacket $\tau_{\text{life}}$
    as an amplitude envelope.
    The autocorrelation time, similar to the particle potential transit time, is estimated as
    $\tau_{ac} = 2\pi |(\zeta - v_g)\Delta k|^{-1}$ as in Section~\ref{subsec:validity_conditions}.
    The quantity $\Delta k$ is the spectral width of the energy-containing oscillations, and can be estimated
    from the saturated spectrum, shown in Fig.~\ref{fig:dns_spectra}(a), as $\Delta k \approx 0.1$.
    The phase velocities range from $\zeta\approx 3$--$5v_t$, so with $v_g\approx 1v_t$
    the autocorrelation frequency is approximately $\omega_\text{ac}=0.4$.

    The importance of wavepacket lifetime can be estimated using the Hilbert transform to obtain the wave envelope $\psi$,
    and computing the group velocity-shifted autocorrelation function $\langle \psi(x,t)\psi(x-v_g\tau,t-\tau)\rangle_L$.
    When this shifted autocorrelation is compared against the unshifted autocorrelation of the envelope during the saturated
    state of the instability a stationary signal is observed.
    The slow evolution of the wavepacket envelope points to the robustness of the saturated state of oscillation.

    Returning to the question of quasilinear applicability, the relaxation time is first estimated from finite differences
    of the Vlasov-Poisson averaged distribution $\langle f\rangle_L$ as $\tau_r^{-1}\approx \mathcal{O}(10^{-3})\omega_p$.
    The maximum growth rate is initially $\omega_i\approx 0.03\omega_p$.
    Thus the inequality $\tau_r^{-1} < \omega_i < \tau_{ac}^{-1}$ is satisfied as $10^{-3} < 10^{-2} < 10^{-1}$.
    The rms bounce frequency of the perturbation is $\omega_b\approx 0.05\omega_p$, but by saturation the bounce frequency has
    grown to $\omega_b\approx 0.1\omega_p$ with maximum values of up to $\omega_b = 0.4\omega_p$ in the largest amplitude
    wavepackets.
    While the instability saturates at only $0.2\%$ of the total thermal energy of the system
    this is enough to make the bounce frequency comparable to the autocorrelation frequency and the quasilinear
    flux is found to be modified~\cite{vlad2004lagrangian} as discussed previously.


    \section{Summary of findings}\label{sec:summary}
    This study revisited the bump-on-tail instability with a
    conservative high-order discontinuous Galerkin method applied to its
    quasilinear theory as well as to a direct simulation of the Vlasov-Poisson equations.
    The initial condition was chosen in the regime of beam parameters for which quasilinear theory should be applicable
    according to Ref.[\onlinecite{thorne2017modern}].
    It was found that quasilinear theory is initially in good agreement with the system evolution but underestimates
    the rate of turbulent momentum flux beginning at field energies halfway to instability saturation.
    The result of the correction beyond the quasilinear approximation is a quenching of the instability in a finite
    time and the replacement of the unstable initial condition by a robust state of oscillation.
    This correction occurs as the phase space eddy turnover time in the largest amplitude wavepackets becomes
    comparable to the phase fluid transit time through the wavepacket potentials.
    Progressive eddy turnover transitions diffusive scattering of resonant electrons by randomly-phased waves to their
    nonlinear phase mixing in potential wells.
    That the wavepackets, or localized concentrations of fluctuation-to-mean
    energy transfer, are responsible for the deviation from quasilinear diffusion can also be understood as the
    breakdown of the quasilinear closure assumption by the development of significant spatial variance to the turbulent
    momentum flux.
    The findings of this article support the theory of phase space density
    granulations~\cite{diamond,dupree1970theory,dupree1972theory} and shed light on the distribution of fluctuations
    with velocity in collisionless plasma turbulence.

    Limitations of quasilinear theory have been known for some time.
    Theories to account for the discrepancies discussed in this article include resonance broadening
    theory and expansion to higher-order in the field amplitude~\cite{krommes2015tutorial,davidson2012methods, diamond}.
    These approaches can considerably complicate the analysis without a necessarily rewarding conclusion, as
    energy conservation in resonance broadening theory is still actively researched~\cite{krommes2015tutorial} and
    higher-order expansions may lead to misleading analysis for nonlinear Landau damping processes~\cite{mouhot2011landau}.
    The findings of this study suggest that using quasilinear theory to predict plasma processes for all but the
    weakest instabilities should be considered as only order-of-magnitude accurate.

    \section{Acknowledgments}
    The authors would like to thank I.A.M.~Datta, W.H.~Thomas, A.~Ho, E.T.~Meier, A.D.~Stepanov, and G.V.~Vogman
    for helpful discussions.
    The information, data, or work presented herein was funded in part by the Air Force Office of Scientific Research
    under award No.~FA9550--15--1--0271.
    This material is also based upon work supported by the National Science Foundation under Grant No.~PHY-2108419.

    \section{Author declarations}
    The authors have no conflicts to disclose.

    \section{Data availability statement}
    The data that support the findings of this study are available from the corresponding author upon reasonable request.


    \bibliographystyle{aipnum4-2}
    \bibliography{main}

\end{document}